\documentclass[prb,a4paper,twocolumn,floatfix,showpacs,showkeys,amsmath,amssymb,nobibnotes,unsortedaddress,superscriptaddress]{revtex4-1}
\usepackage{graphicx}
\usepackage{subfigure}
\usepackage{xcolor}
\usepackage{textcomp}
\usepackage[utf8]{inputenc}
\begin{document}

\title{Impact of Chlorine on the Internal Transition Rates and Excited States of the Thermally Delayed Activated Fluorescence Molecule 3CzClIPN}

\author{Martin Streiter}
\affiliation{Institut für Physik, Technische Universität Chemnitz, 09126 Chemnitz, Germany}
\author{Tillmann G. Fischer}
\affiliation{Institut für Organische Chemie, Universität Leipzig, 04103 Leipzig, Germany}
\author{Christian Wiebeler}
\affiliation{Institut für Analytische Chemie, Universität Leipzig, 04103 Leipzig, Germany}
\affiliation{Leibniz-Institut f\"ur Oberfl\"achenmodifizierung (IOM), 04318 Leipzig, Germany}
\author{Sebastian Reichert}
\affiliation{Institut für Physik, Technische Universität Chemnitz, 09126 Chemnitz, Germany}
\author{Jörn Langenickel}
\affiliation{Zentrum für Mikrotechnologien, Technische Universität Chemnitz, 09126 Chemnitz, Germany}
\author{Kirsten Zeitler}
\affiliation{Institut für Organische Chemie, Universität Leipzig, 04103 Leipzig, Germany}
\author{Carsten Deibel\,*}
\affiliation{Institut für Physik, Technische Universität Chemnitz, 09126 Chemnitz, Germany}
\email{deibel@physik.tu-chemnitz.de}

\begin{abstract}

\noindent \textbf{NOTE:} This document is the Accepted Manuscript version of a Published Work that appeared in final form in The Journal of Physical Chemistry C, copyright American Chemical Society after peer review and technical editing by the publisher. To access the final edited and published work see DOI: https://doi.org/10.1021/acs.jpcc.0c03341

\noindent \textbf{ABSTRACT:} We analyze internal transition rates and the singlet-triplet energy gap of the thermally activated delayed fluorescence (TADF) molecule $\mathrm{3CzClIPN}$, which recently was introduced as an efficient photocatalyst. Distribution and origin of the non-monoexponential decays, which are commonly observed in TADF films, are revealed by analysis of transient fluorescence with an inverse Laplace transform. A numerically robust global rate fit routine, which extracts all relevant TADF parameters by modeling the complete set of data, is introduced. To compare and verify the results, all methods are also applied to the well-known $\mathrm{4CzIPN}$. The influence of the molecular matrix is discussed by embedding low concentrations of TADF molecules in polystyrene films. Finally, quantum chemical calculations are compared to the experimental results to demonstrate that the chlorine atom increases the charge transfer character of the relevant states, resulting in a reduction of the singlet-triplet energy gap.
\end{abstract}

\maketitle

\section{Introduction}
State-of-the-art materials utilized in organic light-emitting diodes (OLED) and photocatalysis (PC) usually contain metal-complexes with rare elements such as iridium and ruthenium.\cite{hung2002recent,reineke2009white,sun2006management,zeitler2009photoredox} Production cost, element scarcity and questions of environmentally friendly mining led to efforts in synthesizing metal-free molecules with comparable and therefore competitive photophysical and chemical properties. In OLED technology, internal quantum efficiencies of 100\,\% can be realized with traditional metal-complexes by triplet recombination (phosphorescence), overcoming the limitation of maximal 25\,\% internal quantum efficiency set by singlet spin statistics.\cite{adachi2001nearly} By converting triplet states into singlet states, so-called TADF (thermally activated delayed fluorescence) materials can---without the necessity of a metal center---equally achieve up to 100\,\% internal quantum efficiency.\cite{uoyama2012highly} In addition to prompt fluorescence emission from the singlet $S_1$ state back into the $S_0$ ground state (referred to as PF in the following), the $T_1$ triplet state (populated by intersystem crossing with rate $k_{\mathrm{ISC}}$) can repopulate the $S_1$ state by reverse intersystem crossing ($k_{\mathrm{RISC}}$), leading to delayed fluorescence (DF). TADF materials are characterized by a donor--acceptor structure which also can be realized by metal-free molecules. Localized wavefunctions on donor and acceptor sites are spatially and energetically well separated, usually by large dihedral angles between donor and acceptor groups. This design leads to a decrease of the energy difference $\Delta E_{\mathrm{ST}}$  between excited $S_1$ and $T_1$ state. These three parameters are theoretically connected by detailed balance in equation (\ref{arr_isc}).\cite{onsager1931reciprocal} The equation suggests that small $\Delta E_{\mathrm{ST}}$ are favored in OLED design because of efficient singlet-from-triplet conversion as well as  important in PC as an indicator of a strong charge transfer state character of the excited state. However, when $\Delta E_{\mathrm{ST}}$ is too low, the exchange integral between ground and excited state approaches zero, leading to low absorption.\cite{hirata2015highly} Therefore, in  both, OLED and PC applications, also parameters such as absorption, nonradiative losses, quantum yield and $k_{\mathrm{RISC}}$ need to be considered.

\begin{figure}[h]
  \centering
\includegraphics*[scale=0.23]{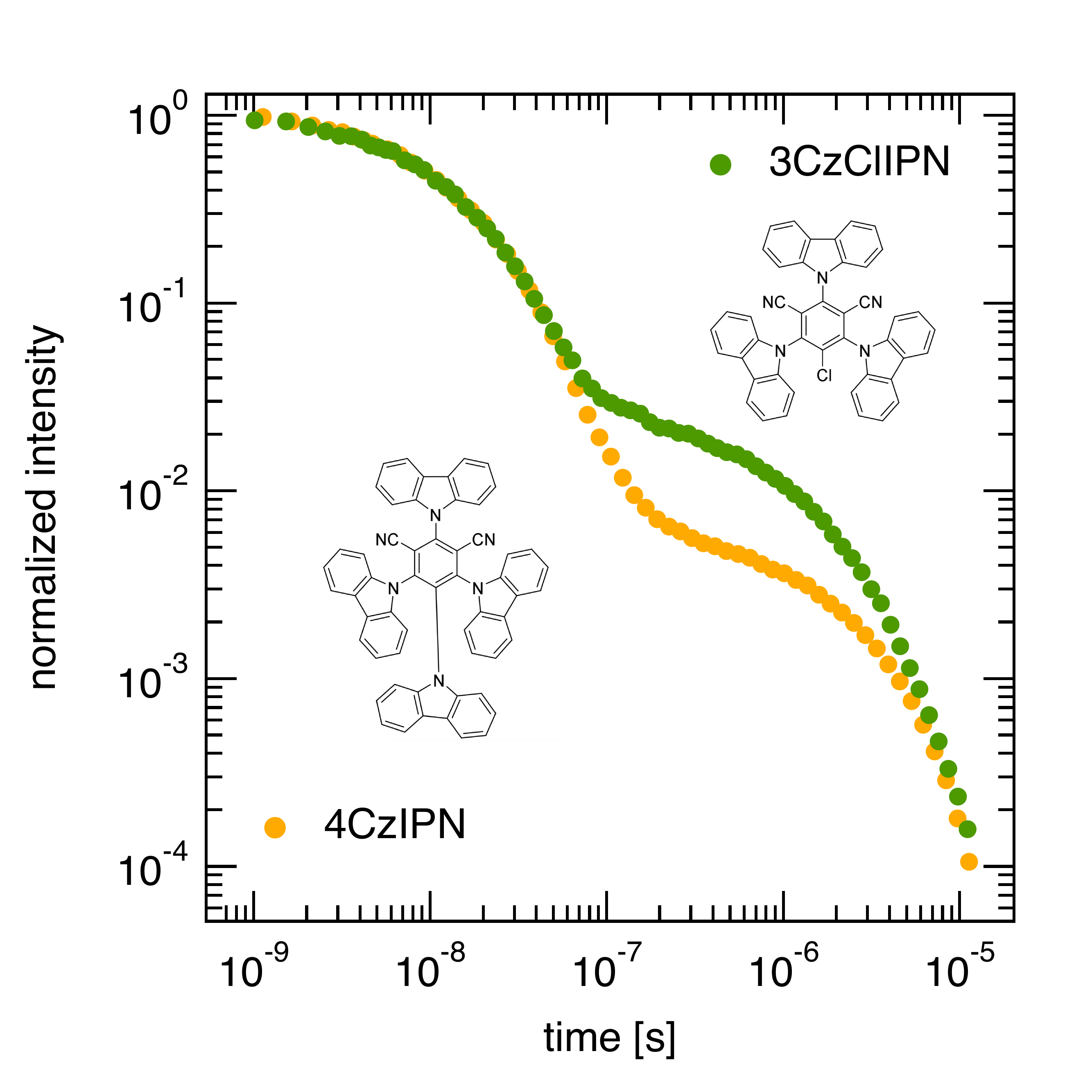}
  \caption{Prompt and delayed fluorescence decay of $\mathrm{4CzIPN}$ and $\mathrm{3CzClIPN}$ films at room temperature.}
    \label{toc}
    \end{figure}

\begin{equation}
k_{\mathrm{RISC}} = k_{\mathrm{ISC}} \exp{\left(-\frac{\Delta E_{\mathrm{ST}}}{k_{\mathrm{B}}T}\right)}.
\label{arr_isc}
\end{equation}

One of the most studied TADF molecule is $\mathrm{4CzIPN}$ (1,2,3,5-tetrakis(carbazol-9-yl)-4,6-dicyanobenzene), which was introduced by Adachi \textit{et al}. \cite{uoyama2012highly} The four carbazole groups act as an electron donor attached to the electron accepting 4,6-dicyanobenzene core at an angle of $60^{\circ}$. Adachi \textit{et al}.\ first demonstrated its efficient properties as a green OLED emitter.\cite{uoyama2012highly}  As a photocatalyst, $\mathrm{4CzIPN}$ shows strong oxidative and reductive  ground state potentials. Recently it was shown that modification of $\mathrm{4CzIPN}$ to $\mathrm{3CzClIPN}$ increases the oxidative ground state potential demonstrating $\mathrm{3CzClIPN}$ as a potential alternative to the well-established, oxidizing  photocatalyst Ir(dF-CF$_3$-ppy)$_2$(dtbbpy)(PF$_6$).\cite{speckmeier2018toolbox,lowry2005single} While the influence of chlorine on TADF molecules is only partly discussed in the literature\cite{xiang2017halogen,kretzschmar2015development}, little is known about the photophysical properties of the photocatalytically potent molecule $\mathrm{3CzClIPN}$. In this paper, we elucidate how chlorine influences the internal transition rates and excited state energies by examining temperature-dependent fluorescence transients with two improved algorithms. Quantum chemical calculations and wavefunction analysis support the experimental results and explain the difference between the two molecules in detail.

\section{Methods}

Photons  emitted via PF and DF are energetically indistinguishable, as they both originate from $S_1$ (see SI). However, in time-dependent measurements, PF and DF  can be separately analyzed as they occur on different timescales with different intensities. Precise and robust routines for measuring, evaluating and simulating TADF parameters are in the focus of current research.\cite{dias2013triplet,dias2017photophysics,zhang2014efficient,gibson2016importance,zhang2012design,hirata2015highly}  Figure \ref{toc} shows the time-dependent fluorescence and chemical structure of the two molecules in focus of this study, $\mathrm{4CzIPN}$ and $\mathrm{3CzClIPN}$. 

We compare two improved evaluation methods based on the work of Dias\cite{dias2017photophysics},  Berberan-Santos\cite{berberan1996unusually,baleizao2007thermally}, Penfold\cite{penfold2018theory}, Scholz\cite{scholz2020investigation}, Haase\cite{haase2018kinetic} and respective co-authors. Both methods require only fluorescence transients without additional measurements of quantum yields, triplet quenching or phosphorescence. They are therefore experimentally straightforward and  sensitive, allowing analysis of highly diluted films with weak emission towards projected single molecule measurements. 

Dias \textit{et al}.\ showed, that $k_{\mathrm{RISC}}$ can be estimated from fluorescence transient measurements when the yields of intersystem crossing and reverse intersystem crossing of the material are high.\cite{dias2017photophysics} More specifically, for a DF/PF ratio larger than $\approx 4$ (which can be determined from the integrals of both regimes in the transient fluorescence signal) and an assumed reverse intersystem crossing yield approaching 100\,\%, $k_{\mathrm{RISC}}$ can be estimated from transient measurements with a DF lifetime $\tau_{\mathrm{DF}}$ by:
 
\begin{equation}
k_{\mathrm{RISC}} =  \frac{1}{\tau_{\mathrm{DF}}} \left( 1+ \frac{\int_{0}^{\infty} DF~\mathrm{d}t}{\int_{0}^{\infty} PF~\mathrm{d}t} \right).
\label{diaseq}
\end{equation}

To determine $\tau_{\mathrm{DF}}$ and the integral DF/PF ratio ($\approx 5$ for 4CzIPN and $\approx 6$ for 3CzClIPN, therefore equation (\ref{diaseq}) was applicable), we analyzed our data with the RegSLapS algorithm which utilizes an inverse Laplace transform.\cite{reichert2019improved} The spectral function $g$ including the effective rates $k_i$ for each component in the multiexponential decay $L$ can be obtained by calculating the inverse Laplace transform of:

\begin{equation}
L(t,T) = \int_0^{\infty}g(k(T))\exp{(-k(T)t)}\mathrm{d}k.
\label{lapla}
\end{equation}

The solution $g$ contains information about the amplitudes $A_i$, effective rates $k_i$ as well as the necessary number of components $i$. Subsequently, every decay can be fitted with the sum of multiple exponential functions (see SI),

\begin{equation}
L(t,T) = \sum_i A_i \exp{(-k_i t)}.
\end{equation}

\begin{figure}[ht]
  \centering
\includegraphics*[scale=0.6]{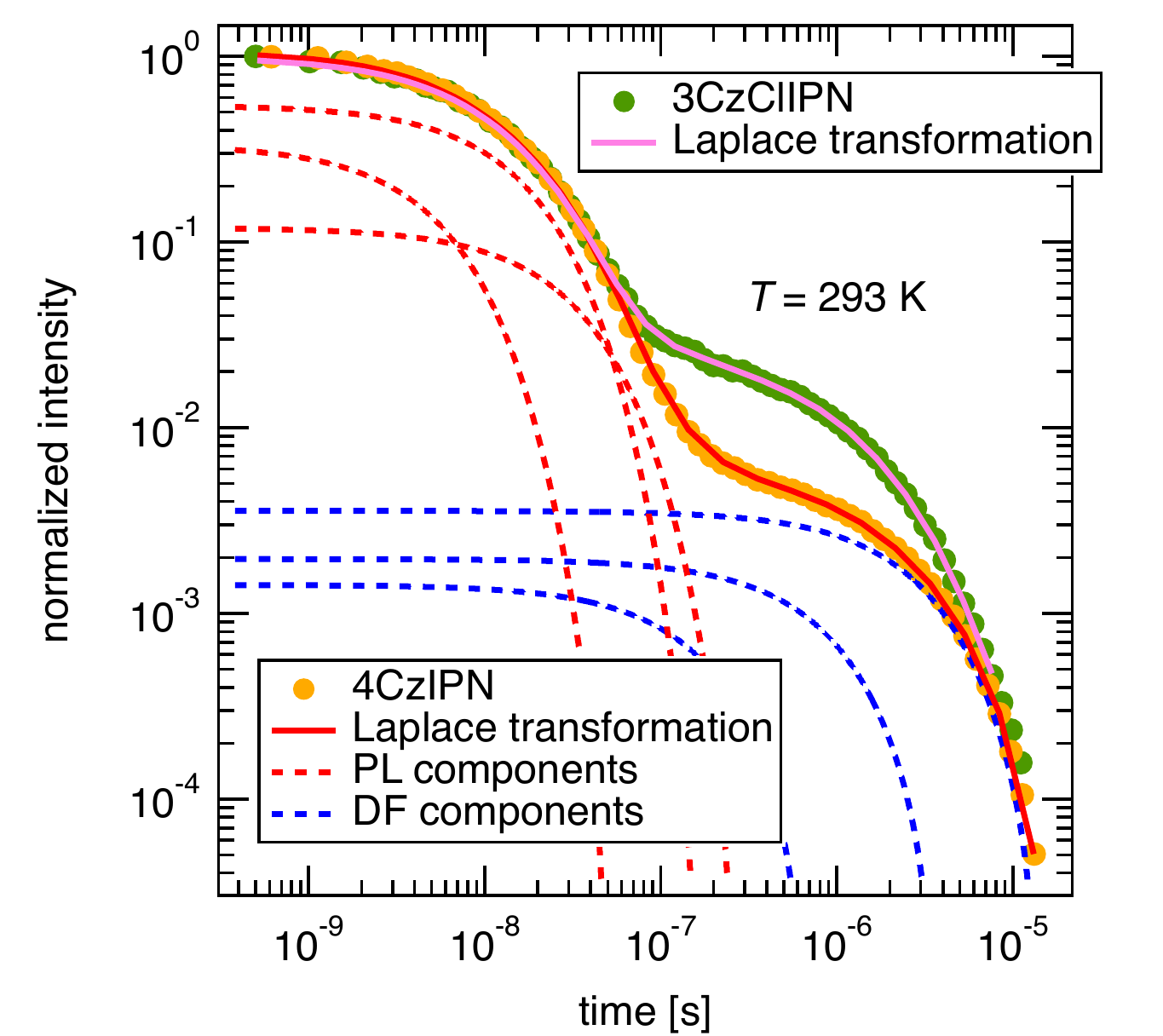}
  \caption{Examplary single components of the inverse Laplace transform algorithm. Resulting fits for neat $\mathrm{4CzIPN}$ and $\mathrm{3CzClIPN}$ films at room temperature.}
    \label{multiexp}
    \end{figure}

This analysis is illustrated in figure \ref{multiexp} for the transient fluorescence of a neat $\mathrm{4CzIPN}$ film and also applied to $\mathrm{3CzClIPN}$ at room temperature. Since solving of equation (\ref{lapla}) is an ill-posed problem, the experimental data has to be almost noise free to avoid instability of the solution $g$. Additionally, uniqueness could not be given as different spectral functions $g$ can have nearly the same Laplace transform. Therefore, we used an extended version of the Tikhonov's regularization as described by Reichert \textit{at al.}\cite{reichert2019improved} to search for the  most stable and reasonable solutions.

With the integral of $A \exp(-kt)$ being $Ak^{-1}$ and an effective DF rate\cite{lakowicz2013principles} (see SI), $k_{\mathrm{RISC}}$ was calculated from the inverse Laplace transform with

\begin{equation}
k_{\mathrm{RISC}} =  \langle k_{\mathrm{DF}}\rangle \left( 1+ \frac{\sum A_i k_{i\mathrm{DF}}^{-1} }{\sum A_j k_{j\mathrm{PF}}^{-1}} \right).
\label{risc_eva}
\end{equation}

Temperature-dependent values of $k_{\mathrm{RISC}}$ were evaluated according to equation (\ref{risc_eva}), and the $k_{\mathrm{RISC}}(T)$ plots (see SI)  were then fitted with

\begin{equation}
k_{\mathrm{RISC}} = k_{\mathrm{A}} \exp{\left(-\frac{\Delta E_{\mathrm{ST}}}{k_{\mathrm{B}}T}\right)}.
\label{arr}
\end{equation}

Note, that the prefactor $k_{\mathrm{A}}$ is described differently in the literature, as discussed later.\cite{onsager1931reciprocal,uoyama2012highly,gibson2016importance,hosokai2019tadf} We refer to this evaluation method as \textit{Laplace fit} throughout the paper. Next, we discuss our second method, the \textit{global rate fit}, which minimizes a global differential rate equation system to the complete, temperature-dependent data set.  Haase \textit{et al}.\ showed that extracting all relevant internal transition rates is possible by fitting the commonly assumed coupled differential equation system (\ref{ode1}--\ref{ode2}) to the data,\cite{haase2018kinetic}

\begin{align}
    \dot S_{1}(t) &= - (k_{\mathrm{F}}+k_{\mathrm{ISC}}) S_{1}(t) + k_{\mathrm{RISC}}T_{1}(t) \label{ode1} \\
    \dot T_{1}(t) &= k_{\mathrm{ISC}}S_{1}(t) - k_{\mathrm{RISC}} T_{1}(t). \label{ode2}
\end{align}

Here, $k_{\mathrm{F}}$ is the sum of radiative and nonradiative depopulation rates of the $S_{1}$ singlet state. The prerequisite for this model is the reduction of the TADF mechanism to a three-level system involving only $S_0$, $S_1$, $T_1$.\cite{penfold2018theory} This ignores the influence of disorder because in an organic molecule film, a broad distribution of states is present. Additionally, recombination from $T_1$ (nonradiative or as phosphorescence) is neglected, as well as a possible contribution from higher excited triplet states.\cite{kobayashi2017contributions} The influence of each of these simplifications will be explained in detail later.

The solution of equation system (\ref{ode1}--\ref{ode2}) equals the sum of two monoexponential functions with different amplitudes and rates, which is insufficient for modeling our data, which deviates from monoexponential decay especially during the PF. From the inverse Laplace transform analysis, we concluded that the non-monoexponential PF decay can be well described with two distinct rates in most cases. These decay characteristics are caused by dynamic and static inhomogeneities of the molecules' conformation and environment in film as discussed later in more detail.\cite{noriega2016uncovering,borner2012efficient}  We assumed that $k_{\mathrm{F}}$ and $k_{\mathrm{ISC}}$ can be treated as temperature-independent parameters because normalized PF decay showed no correlation with temperature. The advantage of fitting a rate equation system to the data, as demonstrated by Haase \textit{et al}., is the direct extraction of all internal rates. However, this evaluation method also consists of two steps: kinetic modeling and fitting of $k_{\mathrm{RISC}}(T)$. We believe that a numerically more robust method is the direct extraction of all relevant parameters from the complete, temperature-dependent data set by global rate fitting in one step. We implemented an algorithm that includes the sum of two singlet states (as predicted by our Laplace evaluation which showed two distinct rate peaks of the fluorescence as shown in the SI). Both singlet states $S_{1,2}$ had a starting population of $N_{1,2}$ after excitation and were depopulated with rates $k_{\mathrm{F^{1,2}}}$, $k_{\mathrm{ISC}}$ and repopulated with $k_{\mathrm{RISC}}$, depending on the triplet population $T_{1,2}(t)$, where $T_{1,2}(0)=0$. 

\begin{align}
    \dot S_{1,2}(t) &= - \left(k_{\mathrm{F^{1,2}}}+k_{\mathrm{ISC}}\right) S_{1,2}(t) + k_{\mathrm{RISC}}T_{1,2}(t)\\
    \dot T_{1,2}(t) &= k_{\mathrm{ISC}}S_{1,2}(t) - k_{\mathrm{RISC}}T_{1,2}(t)\\
    k_{\mathrm{RISC}} &= k_{\mathrm{A}} \exp{\left(-\frac{\Delta E_{\mathrm{ST}}}{k_{\mathrm{B}}T}\right)}
\label{globalfitequ}
\end{align}

\section{Results and discussion}
\begin{figure*}
\centering
\subfigure[$\mathrm{4CzIPN}$, neat film]{\includegraphics*[scale=0.6]{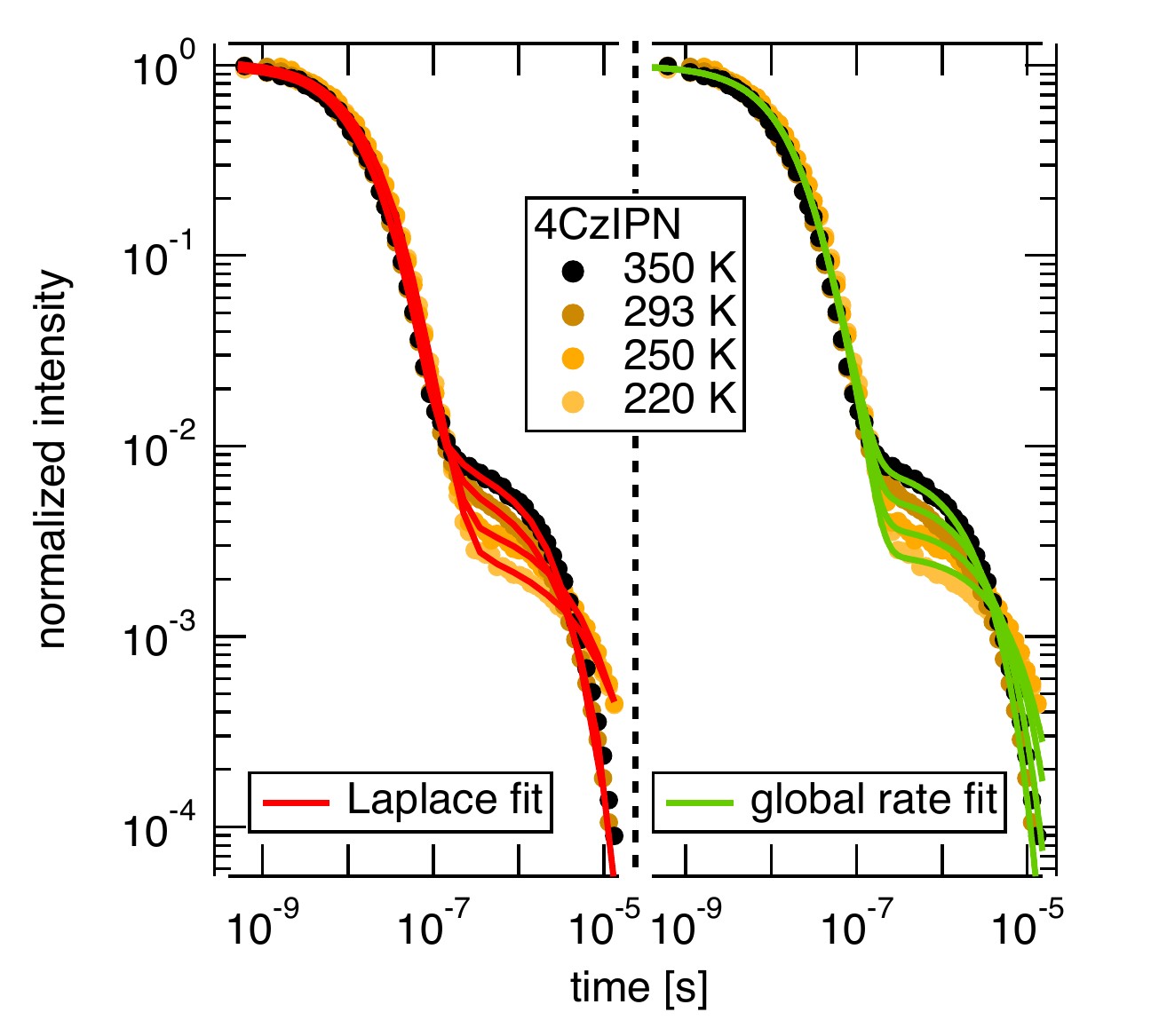}} 
\subfigure[$\mathrm{3CzClIPN}$, neat film]{\includegraphics*[scale=0.6]{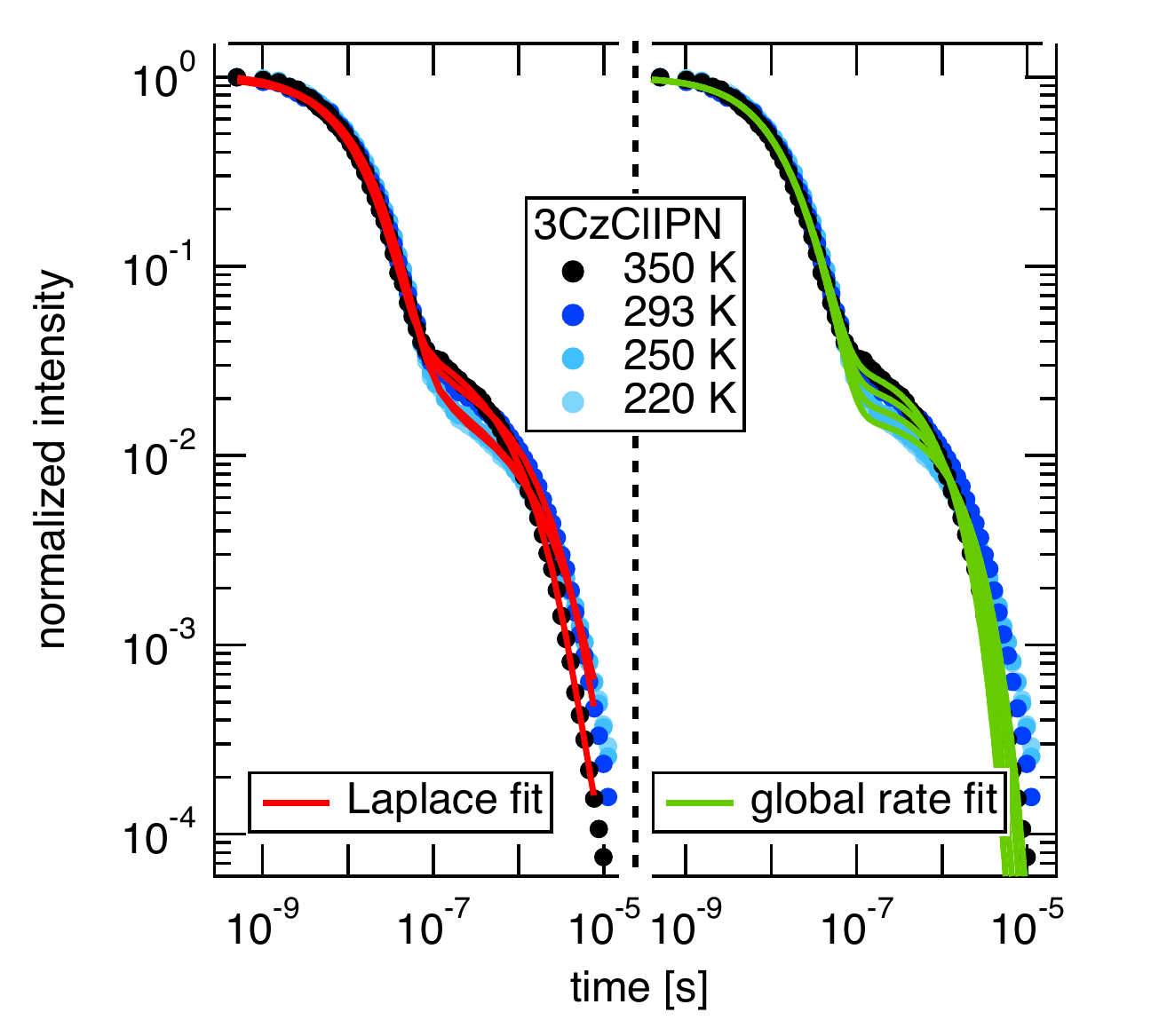}}
\caption{Comparison of both fit methods applied to temperature-dependent transient fluorescence decay of neat TADF films}
\label{fitscl}
\end{figure*}

Figure \ref{fitscl} shows the results of both evaluation methods for neat $\mathrm{4CzIPN}$ and $\mathrm{3CzClIPN}$ films, respectively. Fit results are summarized in table 1. The Laplace fit algorithm fitted all data sets well on the full time scale of the decay while global rate fitting showed small deviations in the DF part of the decay after 1\,\textmu s. We also tested more complex rate equation systems with multiple singlet and triplet states\cite{kobayashi2017contributions} in order to yield precise fits of the DF part, but found the process to be over-parameterized quickly, allowing very good fits but no robust way of extracting the effective values or possible distribution characteristics. This is shown and discussed in detail in the SI.

\begin{table} 
\begin{tabular}{llcccc} 
            &   film                         &$\mathrm{4CzIPN}$ & &$\mathrm{3CzClIPN}$ & \\
    method          &                          &neat &in PS  &neat &in PS \\
\hline
\hline
\textbf{Laplace}&\textbf{fit}      &&&&\\
$\Delta E_{\mathrm{ST}}$&$[\mathrm{meV}]$&54 &65 &34 &38 \\
$k_{\mathrm{RISC}}$&$[\mathrm{10^6\,s^{-1}}]$  &0.6& 0.7&2.1&1.8\\
$k_{\mathrm{A}}$&$[\mathrm{10^6\,s^{-1}}]$   &5.1 & 8.7& 7.6 & 8.9\\
$\langle \tau_{\mathrm{PF}} \rangle$&$[\mathrm{ns}]$   &15 & 14& 16 & 12\\
\hline
\textbf{Global}&\textbf{rate fit}      &&&&\\
$\Delta E_{\mathrm{ST}}$&$[\mathrm{meV}]$&54 &62 &33 &30\\
 $\langle k_{\mathrm{F}} \rangle $&$[\mathrm{10^6\,s^{-1}}] $  &14&18&17&12\\
$k_{\mathrm{RISC}}$&$[\mathrm{10^6\,s^{-1}}]$  &0.7&0.8&3.9&2.6\\
$k_{\mathrm{A}}$&$[\mathrm{10^6\,s^{-1}}]$   &6.2 &9.4 &11 &8.5\\
$k_{\mathrm{ISC}}$&$[\mathrm{10^6\,s^{-1}}]$   &16&21&31&37\\
\hline
\hline
PLQY &$[\%]$  & n/a & 99 & n/a & 54 \\

\end{tabular}

\caption{Experimental results, $k_{\mathrm{RISC}}$, $\langle k_{\mathrm{F}}\rangle$ and $\langle \tau_{\mathrm{PF}} \rangle$ at 293\,K. PLQY for neat films was not measured due to strong reabsorption.}
\label{ergebnisse}
\end{table}

We now discuss the results for neat films and compare both evaluation methods with our calculated values and the literature. Both evaluation methods delivered  matching values $\Delta E_{\mathrm{ST}} = (54\pm 5)$\,meV for a neat $\mathrm{4CzIPN}$ film. Depending on film matrix (or solvent), but also measurement technique and evaluation method, values between 30\,meV and 140\,meV are reported for $\mathrm{4CzIPN}$ in the literature.\cite{yurash2019photoluminescence,uoyama2012highly,kobayashi2017contributions,olivier2017nature,noda2019critical,menke2016exciton,niwa2014temperature,ishimatsu2013solvent} However, in neat film with a similar experiment, Olivier \textit{et al}.\ measured 42\,meV (and simulated value: 60\,meV), which is close to our result.\cite{olivier2017nature}  In $\mathrm{3CzClIPN}$, we found $\Delta E_{\mathrm{ST}} = (33\pm 8)$\,meV. Note, that this is the first  value of $\Delta E_{\mathrm{ST}}$ reported for $\mathrm{3CzClIPN}$. The error in determining $\Delta E_{\mathrm{ST}}$ increases for low $\Delta E_{\mathrm{ST}}$ values because the temperature dependence of DF becomes small compared to the dynamic range of the complete decay. The mechanism behind the chlorine-induced decrease of $\Delta E_{\mathrm{ST}}$ is explained later in the molecular orbital calculation part. As explained in the introduction, internal transition rates are connected to the $S_1$ and $T_1$ energies as well as to their relative position. In the literature, $k_{\mathrm{ISC}}$ values between $(1-7)\cdot 10^{7}\,\mathrm{s^{-1}}$ are reported.\cite{olivier2017nature,kobayashi2019intersystem,einzinger2017shorter,noda2019critical,kobayashi2017contributions,uoyama2012highly,yurash2019photoluminescence} At room temperature, $k_{\mathrm{RISC}}$ values between $(6-12)\cdot 10^{5}\,\mathrm{s^{-1}}$ are reported.\cite{noda2019critical,einzinger2017shorter,kim2017concentration,olivier2017nature,uoyama2012highly} Our results lie between these values (table \ref{ergebnisse}). With the global rate fit method, it is possible to simultaneously determine $k_{\mathrm{A}}$ and  $k_{\mathrm{ISC}}$. This is enabled by globally fitting for the optimized and temperature-independent $k_{\mathrm{ISC}}$. Within our temperature range between 220\,K and 350\,K, we assume  $k_{\mathrm{ISC}}$ to be constant.\cite{kobayashi2019intersystem} As shown in table 1, the prefactor $k_{\mathrm{A}}$ was extracted with very good consistency by both evaluation methods. However, the global rate fit revealed that the actual $k_{\mathrm{ISC}}$ is higher. Such results can also be found in the literature, when comparing published $k_{\mathrm{ISC}}$ values to the actual prefactor in the according $k_{\mathrm{RISC}}(T)$ plots. The difference between $k_{\mathrm{ISC}}$ and $k_{\mathrm{A}}$ can be interpreted as intramolecular transition pathways which deviate from detailed balance, such as nonradiative losses\cite{olivier2017nature},  spin-orbit coupling\cite{hosokai2019tadf}, exciton diffusion\cite{yurash2019photoluminescence} and dynamic asymmetries which result from internal reorganization of the excited charge transfer state. The absolute photoluminescence quantum yield (PLQY) of fluorophores (given by the ratio of emitted/absorbed photons, see SI for details) is dependent on a variety of parameters, such as molecular environment. To avoid reabsorption and fluorophore-fluorophore interaction, we measured the PLQY by embedding 1\,wt\% of TADF emitters in a PS matrix, a technique which is known from single molecule spectroscopy.  Note, that the spectral characteristics of both, photoluminescence and electroluminescence spectra of $\mathrm{4CzIPN}$ and $\mathrm{3CzClIPN}$ are similar, as shown in the SI. We measured a PLQY of ($99\pm1$)\,\% for $\mathrm{4CzIPN}$ and ($54\pm5$)\,\% for $\mathrm{3CzClIPN}$.  Adachi \textit{et al.} demonstrated that a reduction of the number of carbazole groups and change of substituents can decrease PLQY.\cite{uoyama2012highly}  The increased spin-orbit coupling caused by chlorine\cite{xiang2017halogen} may additionally introduce nonradiative transition pathways, thus lowering the PLQY. To examine how nonradiative depopulation of $T_1$ would affect the transients, we performed a global rate fit by adding a nonradiative pathway $T_1 \rightarrow S_0$ with rate ($k^T_{\mathrm{nr}}$):

\begin{equation}
\dot T_1(t) = k_{\mathrm{ISC}}S_1(t) - k_{\mathrm{RISC}}T_1(t) - k^T_{\mathrm{nr}}T_1(t).
\end{equation}

This approach showed that high $k^T_{\mathrm{nr}}$ rates would lead to a faster DF decay than we experimentally observed. Therefore, $S_1 \rightarrow S_0$ is probably the main nonradiative transition channel in $\mathrm{3CzClIPN}$. Compared to $\mathrm{4CzIPN}$, we also found a reduced PLQY of $\mathrm{3CzClIPN}$ in solution  (see SI).

% found a reduced electroluminescence signal of $\mathrm{3CzClIPN}$ compared to $\mathrm{4CzIPN}$, when embedding the emitters in a contacted device,

%As shown in figure \ref{toc}, the PF decay of both molecules is similar. From the 99\,\% PLQY of $\mathrm{4CzIPN}$ follows that $k_{\mathrm{F}} = k_{\mathrm{rad}}$ and therefore  $\tau^{-1}_{\mathrm{PF}} = k_{\mathrm{rad}} + k_{\mathrm{ISC}}$.
%\begin{equation}
%    \dot T_1(t) = k_{\mathrm{ISC}}S_1(t) - k_{\mathrm{RISC}}T_1(t) - k^T_{\mathrm{nr}}T_1(t)
%\end{equation}
%PLQY: 3x more contributions from recovered triplet states in 3Cl, therefore stronger influenced by potential nonradiative losses of T1. fewer carbazol groups metioned by adachi2012, chlorine may increase spin-orbit coupling\cite{xiang2017halogen,kretzschmar2015development} but may increase nonrad singlet losses which are absent in 4cz. stronger deviation between ka and kisc in 3cl in our global rate fit indicates difference (2x compared to 4x)
%which indicates that nonradiative losses from $T_1$ may be considered in equation (\ref{globalfitequ}) in the future. --> model mit losses testen und einfach sagen dass hier halt die nonloss annahme gemacht wurde
% we made test with nonrad rate but this shortens the lifetime of the TADF decay and is therefore wrong, more likely losses are from S1

%%%%%%%%%%%%%%%%%%%%%%%% Quantum Chemical Calculations %%%%%%%%%%%%%%%%%%%%%

To understand the molecular origin of the difference in $\Delta E_{\mathrm{ST}}$  between $\mathrm{4CzIPN}$ and $\mathrm{3CzClIPN}$, we performed quantum chemical calculations either based on density functional theory (DFT) or with Post-Hartree-Fock methods. A more detailed discussion of our calculations and in particular the results from regular time-dependent DFT (TD-DFT) calculations with and without Tamm-Dancoff approximation (TDA) can be found in the SI. Here, the focus of the presentation will be on the results from simplified TD-DFT (sTD-DFT) and simplified TDA (sTDA), which will be assessed \textit{via} comparison with results from approximate coupled cluster singles and doubles (CC2) calculations with and without spin-component scaling (SCS), see table \ref{sim_ST}.

\begin{table} [htbp]
\begin{tabular}{lcllcll} 

		&	$\mathrm{4CzIPN}$ &&&  	$\mathrm{3CzClIPN}$ & & \\

	&	$\Delta E_{\mathrm{ST}}$ & $S_1$ & $T_1$  & $\Delta E_{\mathrm{ST}}$ &$S_1$ & $T_1$ \\
	method &	$[\mathrm{meV}]$ & $[\mathrm{eV}]$ & $[\mathrm{eV}]$ &$[\mathrm{meV}]$ & $[\mathrm{eV}]$ & $[\mathrm{eV}]$ \\
\hline
\hline
\textbf{B3LYP} 	&&&&&&\\
sTD-DFT 		&46   & 2.411 & 2.365        &41   & 2.341 & 2.300\\
sTDA 		&56   & 2.421 & 2.365        &47   & 2.347 & 2.300\\
\hline
\textbf{CAM-B3LYP} &&&&&&\\
sTD-DFT 		&77   & 3.056 & 2.979         &49   & 3.013 & 2.964\\
sTDA 		&87   & 3.067 & 2.980         &54   & 3.019 & 2.965\\
\hline
\textbf{Post-HF} &&&&&&\\
CC2                 &43 &2.842   &2.798         &22   & 2.847 & 2.824\\
SCS-CC2          &38 &3.179  &3.140         &16   & 3.204 & 3.188
\end{tabular}

\caption{Vertical singlet-triplet gaps and energies of the first excited singlet and triplet states relative to the optimized ground state.}
\label{sim_ST}
\end{table}

The central quantity in our analysis is the energy difference between the $S_1$ and $T_1$ states for the ground state optimized geometries, \textit{i.e.} the vertical $\Delta E_{\mathrm{ST}}$. For a more rigorous comparison with experiments, the energy difference of  these two states in the corresponding minima should be determined, \textit{i.e.} the adiabatic $\Delta E_{\mathrm{ST}}$.\cite{penfold2015} Given the similarity of the two states for $\mathrm{4CzIPN}$ and $\mathrm{3CzClIPN}$, we expect that the adiabatic singlet-triplet gap can be approximated by its vertical counterpart, which has been proposed by Penfold as well as by Tian \textit{et al}.\ \cite{penfold2015, tian2016} This is further corroborated by two theoretical studies of TADF molecules that reported similar trends for both definitions of $\Delta E_{\mathrm{ST}}$.\cite{sun15, olivier2017} 

The unscaled CC2 calculations yield values for the vertical $\Delta E_{\mathrm{ST}}$ that underestimate the experimental ones by \textit{ca.} \(10 \, \mathrm{meV}\). Nonetheless, the difference of this energy between the two compounds of \(22 \, \mathrm{meV}\) is close to its experimental counterpart. The application of SCS results in further lowering of both energies by around \(5 \, \mathrm{meV}\), so the difference remains nearly constant. Overall, unscaled CC2 calculations to determine vertical $\Delta E_{\mathrm{ST}}$ yield the best match with experiment among the employed methods (see the SI for a complete overview).

In case of the sTD-DFT/sTDA calculations, the results for individual $\Delta E_{\mathrm{ST}}$ from sTD-DFT based on B3LYP ground state calculations are the closest to the CC2 reference. However, this property is only overestimated by \(5 \, \mathrm{meV}\) for $\mathrm{4CzIPN}$, whereas for $\mathrm{3CzClIPN}$ the overestimation is \(19 \, \mathrm{meV}\). Therefore, the difference in $\Delta E_{\mathrm{ST}}$ is much too low with \(5 \, \mathrm{meV}\). Employing sTDA leads to an increase of the \(S_1\) energies, but hardly affects the \(T_1\) energies. Owing to this, $\Delta E_{\mathrm{ST}}$ increases and also their difference becomes slightly larger with \(9 \, \mathrm{meV}\). Analyzing the results from sTD-DFT and sTDA calculations of $\Delta E_{\mathrm{ST}}$ based on ground state calculations with the range-separated hybrid functional CAM-B3LYP, similar trends are found. However, the values for $\Delta E_{\mathrm{ST}}$ are even higher than their B3LYP counterparts. Nonetheless, the sTD-DFT calculations with CAM-B3LYP functional yield a difference of \(28 \, \mathrm{meV}\) for $\Delta E_{\mathrm{ST}}$ between the two compounds, which is the closest to the CC2 reference and experiment from all DFT-based calculations. Therefore, this approach appears promising for the investigation of trends in this property, whereas sTD-DFT calculations based on B3LYP ground state calculations might underestimate such changes. Overall, the semiempirical sTD-DFT and sTDA methods yield values for $\Delta E_{\mathrm{ST}}$ that are much closer to the Post-Hartree-Fock calculations than the ones from regular TD-DFT and TDA calculations (see SI).

\begin{figure*}[htbp]
\centering
%\subfigure[$\mathrm{4CzIPN}$]{\includegraphics*[scale=0.5]{$\mathrm{4CzIPN}$_NTOs.png}} 
%\hspace{2cm}
%\subfigure[$\mathrm{3CzClIPN}$]{\includegraphics*[scale=0.5]{$\mathrm{3CzClIPN}$_NTOs.png}}
\includegraphics[width= 0.8 \linewidth]{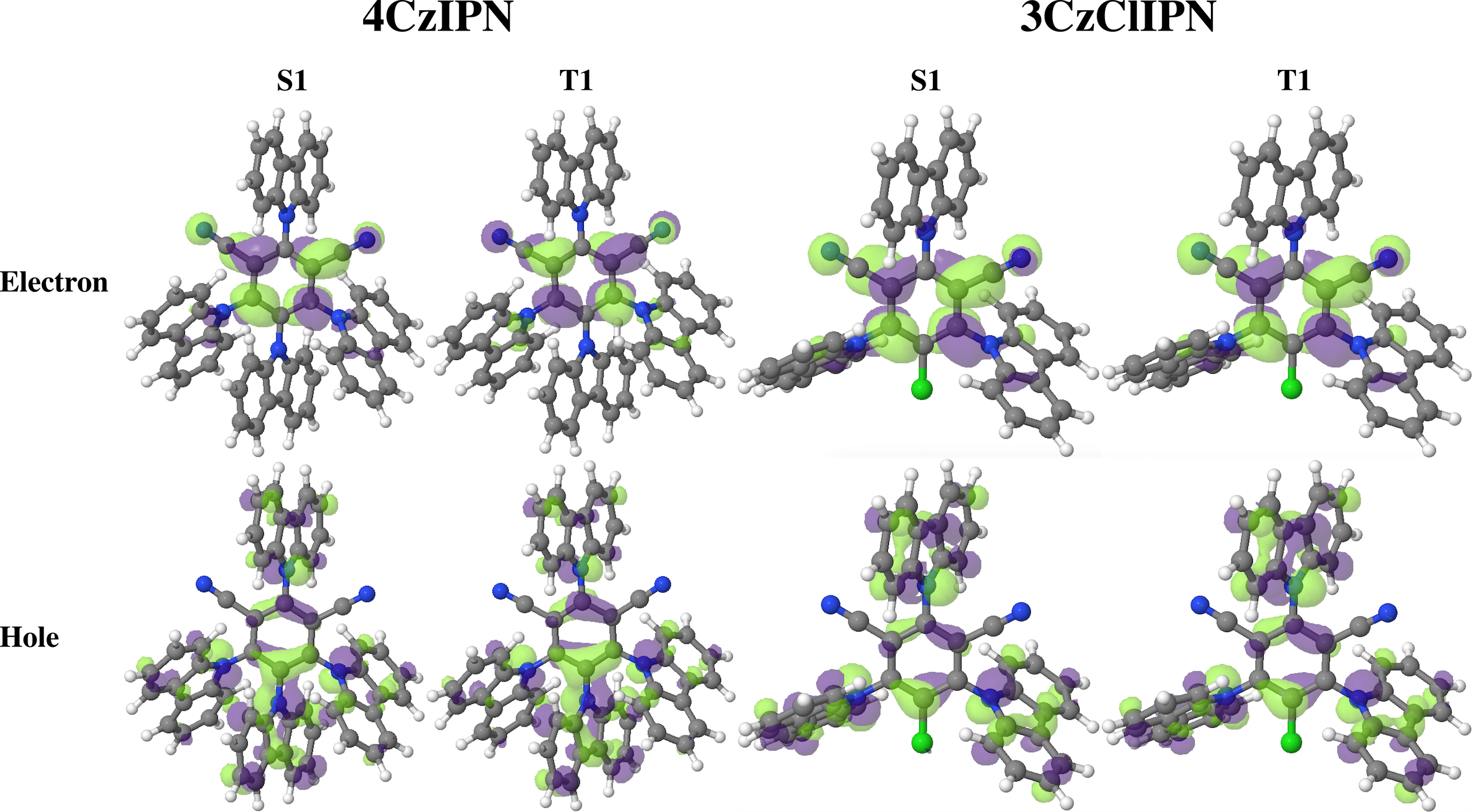}
\caption{Natural transition orbitals (NTO) of the first excited singlet state ($S_{1}$) and the first triplet state ($T_{1}$) from the CC2 calculations. For comparability, all NTOs are visualized with the same isovalue.}
\label{orbitals}
\end{figure*}

Insights into the nature of the \(S_{1}\) and \(T_{1}\) states can be obtained by visualizing the natural transition orbitals (NTOs), see figure \ref{orbitals}. NTOs are compact orbital representations for the transition density resulting in a minimum number of electron-hole excitations.\cite{Martin2003} For all investigated electronic transitions, one pair of NTOs is dominant. These orbitals demonstrate that the \(S_{1}\) and \(T_{1}\) states are rather similar and there are also only small differences between the two compounds. The excited electron is mainly localized at the two cyano groups and the central benzene moiety. The hole is also found at the latter part, but it is also present at the carbazole units. It appears that the charge transfer character is more pronounced for $\mathrm{3CzClIPN}$ than for $\mathrm{4CzIPN}$.

To quantifiy these findings, we performed wavefunction analysis and report the outcome for unscaled CC2 in the following. For this purpose, the molecules are divided into three fragments: the first one consists of the carbazole units, the second one of the benzene unit plus the chlorine atom in case of $\mathrm{3CzClIPN}$, and the third one of the cyano groups. Based on this fragmentation, we obtain similar values from wavefunction analysis for the \(S_{1}\) and \(T_{1}\) states of each compound. The charge transfer character can be used as a quantitative descriptor. It ranges from 0 for an excitation completely localized at one fragment to 1 for the case that electron and hole are localized at different fragments.\cite{Plasser2012} For $\mathrm{4CzIPN}$, this charge transfer character is \(0.78\) and \(0.77\) for \(S_{1}\) and \(T_{1}\), respectively, and it increases to \(0.84\) and \(0.83\) for $\mathrm{3CzClIPN}$. Therefore, the increase in charge transfer character leads to a decrease of $\Delta E_{\mathrm{ST}}$  and further information on the performed wavefunction analysis along with the results for $\mathrm{3CzHIPN}$ is given in the SI.

%%%%%%%%%%%%%%%%%%% End Quantum Chemical Calculations%%%%%%%%%%%%%%%%%

Next, we discuss the influence of the molecular environment in our measurements. Molecules which are sufficiently separated are known to show differences and fluctuations of their emission\cite{ambrose1991fluorescence,krause2011spectral} and absorption\cite{streiter2016dynamics} spectra and lifetime.\cite{borner2012efficient} This is caused by static differences of the molecular environment and dynamic variations of, for example, the side group movement and alignment of the molecule. The influence of the matrix (host) and molecular arrangement on TADF properties is widely discussed in the literature.\cite{northey2017role,cho2014high,mamada2017highly,hasegawa2018well,cucchi2019influence,takeda2019fluorescence,olivier2017nature,kim2017concentration,hosokai2019tadf,ishimatsu2013solvent} To study the non-monoexponential decay characteristics of $\mathrm{4CzIPN}$ and $\mathrm{3CzClIPN}$ in film, we compared neat films to PS films doped with a low (1\,wt\%) concentration of TADF molecules. Neat films of $\mathrm{4CzIPN}$ and $\mathrm{3CzClIPN}$ show a strong red-shift and lower PLQY in comparison to doped PS films (see SI). This is mainly caused by selfabsorption (reabsorption within the film) and self-quenching.\cite{ahn2007self,kim2017concentration} Polystyrene films doped with 1\,wt\% of $\mathrm{4CzIPN}$ showed a slightly increased $\Delta E_{\mathrm{ST}}$ compared to neat films. Considering the error of $\pm 8\,$meV, $\mathrm{3CzClIPN}$ shows similar results as neat film or embedded in PS. This indicates, that solid state solvation effects caused by matrix polarization and static charge transfer state stabilization\cite{northey2017role} are minor when embedding $\mathrm{4CzIPN}$ in PS, and negligible in case of $\mathrm{3CzClIPN}$. This result is in agreement with a supplementary measurement by Olivier \textit{et al}.\, who found similar decay characteristics of a neat 2CzPN (1,2-bis(carbazol-9-yl)-4,5-dicyanobenzene) film compared to a doped 98\,wt\% PS film.\cite{olivier2017nature} We conclude that---although spectral properties are strongly affected by the matrix---the similar stretched exponential characteristics of the transients are not only caused by static or dynamic film disorder, but by the molecules' large dihedral side group angles. This is in agreement with the work of Hasegawa \textit{et al}.\ who found that the electronic states of $\mathrm{4CzIPN}$ films are similar to the monomer.\cite{hasegawa2018well}

\section{Conclusions}
By investigating steady-state and time-dependent fluorescence characteristics, we elucidated the role of chlorine on the internal transition rates and singlet-triplet energy splitting $\Delta E_{\mathrm{ST}}$ of $\mathrm{3CzClIPN}$ films compared to $\mathrm{4CzIPN}$. Although the photoluminescence and electroluminescence spectra of both materials is similar, quantum yield and temperature-dependent delayed fluorescence are drastically influenced by the chlorine group. Photoluminescence quantum yield was reduced by factor 2 in $\mathrm{3CzClIPN}$, due to nonradiative transitions from the excited singlet to the ground state. Reverse intersystem crossing is strongly increased in $\mathrm{3CzClIPN}$ by factor 3--5 due to the lowered $\Delta E_{\mathrm{ST}}$. Quantum chemical calculations explained these findings by showing that chlorine increases the charge transfer character of the relevant states. This may also explain the increased photocatalytic efficiency of $\mathrm{3CzClIPN}$ in oxidations which was recently demonstrated.\cite{speckmeier2018toolbox} For potential applications of $\mathrm{3CzClIPN}$ in OLED devices, the nonradiative losses are a disadvantage.  Our findings underline that precise knowledge of the excited states alone is insufficient in predicting the performance of a TADF molecule for potential applications in photocatalysis and OLED devices. To gain the necessary additional information about internal transition rates, we introduced a robust global evaluation method for transient fluorescence data which outputs all parameters of the specified TADF model in one single optimization step. By comparing neat films to diluted polystyrene films, we concluded that the often observed stretched exponential transient characteristics of TADF materials are a feature of large dihedral side group angles in condensed phase. 

\section{Methods}
All decay data sets and fit functions were normalized in order to reduce optimization parameters. The algorithm then minimized the global difference between temperature-dependent data and the solution of the temperature-dependent rate equations. The global difference vector was weighted with a three part step function, emphasizing the $1-5$\,\textmu s part of the decay, where DF is measured with the best signal-to-noise ratio. All fit parameters ($N_1$, $N_2$, $k^1_{\mathrm{F}}$, $k^2_{\mathrm{F}}$, $k_{\mathrm{ISC}}$, $k_{\mathrm{A}}$, $\Delta E_{\mathrm{ST}}$) are global.  TADF and TADF:polystyrene (PS) films were prepared by spincoating. Silicon substrates with 100\,nm of thermally grown oxide were used. Substrates and glassware were cleaned by annealing at $450^{\circ}\,$C for two hours in a laboratory oven. PS with a molar mass of $20\,\mathrm{kg\,mol^{-1}}$ was doped with $1\,\mathrm{wt\%}$ of $\mathrm{4CzIPN}$ and $\mathrm{3CzClIPN}$, respectively, at a total concentration of $10\,\mathrm{mg\,ml^{-1}}$. For neat dye films, $0.1\,\mathrm{mg\,ml^{-1}}$ of dye in toluene was used.  The samples were measured with a homebuilt laser scan confocal microscope in a  cryostate (Janis, ST500) with a Zeiss LWD 63$\times$, NA=0.75 objective. Fluorescence decay was acquired with time-correlated single photon counting (TCSPC) using a Perkin Elmer avalanche photodiode (APD) and a PicoHarp300 TCSPC module (PicoQuant). The PicoHarp300 was set to the maximum time window of 33.55\,\textmu s with a time resolution of 0.512\,ns per channel. Data was logarithmically binned and the background level substracted. The samples were excited with a 465\,nm pulsed diode laser (PicoQuant, $\tau_{\mathrm{IRF}} < 1\,$ns) at a repetition rate of 25\,kHz. Fluorescence was detected with a 500\,nm longpass filter. Each decay curve was measured by integrating for 15\,minutes. The dark count rate of the APD was 50\,cts (counts per second). The objective was defocussed until the fluorescence signal was below 1000\,cts  on the APD to ensure falling below the general 5\,\% limit of the  TCSPC technique (signal photons to laser pulses per time interval).\cite{picoquant} This condition avoids pile-up effects caused by the total dead time of the TCSPC system. Although the dark count to signal ratio suggests a signal to noise ratio of only 10:1, the effective ratio is better than 10000:1 because 1000\,cts of signal  refer to the decay occurring within few microseconds while the dark count rate refers to one second.  We want to emphasize that measuring time-dependent fluorescence in both, nanosecond and microsecond (or millisecond) time regimes with equal experimental parameters is challenging and demands ensuring appropriate equilibrium conditions of optical excitation and decay. Measuring at non-equilibrium conditions  leads to wrong results and conclusions. This is especially the case at low temperatures, materials with long DF lifetimes and when measuring PF and DF separately with different excitation sources (or laser repetition rates) on different time scales. 
% power 10uW?

\begin{acknowledgements}
% We thank Fraunhofer ENAS for contributing to OLED production. 
C. W. acknowledges funding by the German Research Foundation (DFG) via a return fellowship (reference number: WI 4853/2-1). The quantum chemical calculations were performed on ressources provided by the Leipzig University Computing Centre and by the Paderborn Center for Parallel Computing. C. W. also thanks Stefan Zahn, Bernd Abel, and J\"org Matysik for discussions and support.
\end{acknowledgements}

%\appendix

\section*{Corresponding Author}
Email: deibel@physik.tu-chemnitz.de

\section*{Notes}
The authors declare no competing financial interest.

\section*{Supplementary Information}
Inverse Laplace fit results, additional rate model fit results, global rate fit results, $k_{\mathrm{RISC}}(T)$ plots, simulation results, UV-Vis spectra, quantum yield, streak measurements, quantum chemical calculations

\newpage

\normalsize{\textbf{REFERENCES}}
\section*{~}

%\bibliographystyle{ieeetr}
%\bibliography{A_bibo}    

\providecommand{\latin}[1]{#1}
\makeatletter
\providecommand{\doi}
  {\begingroup\let\do\@makeother\dospecials
  \catcode`\{=1 \catcode`\}=2 \doi@aux}
\providecommand{\doi@aux}[1]{\endgroup\texttt{#1}}
\makeatother
\providecommand*\mcitethebibliography{\thebibliography}
\csname @ifundefined\endcsname{endmcitethebibliography}
  {\let\endmcitethebibliography\endthebibliography}{}

%%%%%%%%%%%%%%%%%%%%%%%%%
% \clearpage
% \section*{TOC graphic}

% \begin{figure}[h]
%   \centering
% \includegraphics*[scale=0.27]{compare.pdf}

%     \end{figure}
    
% \clearpage

\end{document}

\title{Supporting Information\\ Impact of chlorine on the internal transition rates and excited states of the thermally delayed activated fluorescence molecule 3CzClIPN}

\author{Martin Streiter}
\affiliation{Institut für Physik, Technische Universität Chemnitz, 09126 Chemnitz, Germany}
\author{Tillmann Fischer}
\affiliation{Institut für Organische Chemie, Universität Leipzig, 04103 Leipzig, Germany}
\author{Christian Wiebeler}
\affiliation{Institut für Analytische Chemie, Universität Leipzig, 04103 Leipzig, Germany}
\affiliation{Leibniz-Institut f\"ur Oberfl\"achenmodifizierung (IOM), 04318 Leipzig, Germany}
\author{Sebastian Reichert}
\affiliation{Institut für Physik, Technische Universität Chemnitz, 09126 Chemnitz, Germany}
\author{Jörn Langenickel}
\affiliation{Zentrum für Mikrotechnologien, Technische Universität Chemnitz, 09126 Chemnitz, Germany}
\author{Kirsten Zeitler}
\affiliation{Institut für Organische Chemie, Universität Leipzig, 04103 Leipzig, Germany}
\author{Carsten Deibel}
\affiliation{Institut für Physik, Technische Universität Chemnitz, 09126 Chemnitz, Germany}
\email{deibel@physik.tu-chemnitz.de}

\maketitle

\clearpage
\section{Equations}

\noindent Effective rate $k_{\mathrm{DF}}$:
\begin{equation}
\langle k_{\mathrm{DF}} \rangle = \frac{A_1 k_{\mathrm{DF1}}^{-1} + A_2 k_{\mathrm{DF1}}^{-1}}{A_1 k_{\mathrm{DF1}}^{-2} + A_2 k_{\mathrm{DF2}}^{-2}}  
\label{effk1}
\end{equation}

\noindent Effective photoluminescence rate determined with amplitude-weighted rates $k_{\mathrm{F}}$:
\begin{equation}
\langle k_{\mathrm{F}} \rangle = \frac{A_1 k_{\mathrm{F1}}^{-1} + A_2 k_{\mathrm{F1}}^{-1}}{A_1 k_{\mathrm{F1}}^{-2} + A_2 k_{\mathrm{F2}}^{-2}} 
\label{effk2}
\end{equation}

\noindent Effective photoluminescence lifetime determined with the two main amplitude-weighted exponential rates from inverse Laplace transform:
\begin{equation}
\langle \tau_{\mathrm{PL}} \rangle = \frac{A_1 k_{\mathrm{F1}}^{-2} + A_2 k_{\mathrm{F2}}^{-2}}{A_1 k_{\mathrm{F1}}^{-1} + A_2 k_{\mathrm{F1}}^{-1}} 
\label{effk3}
\end{equation}

\section{Laplace transform fit}

\begin{figure}[h]
\centering
\subfigure[$\mathrm{4CzIPN}$]{\includegraphics*[scale=0.45]{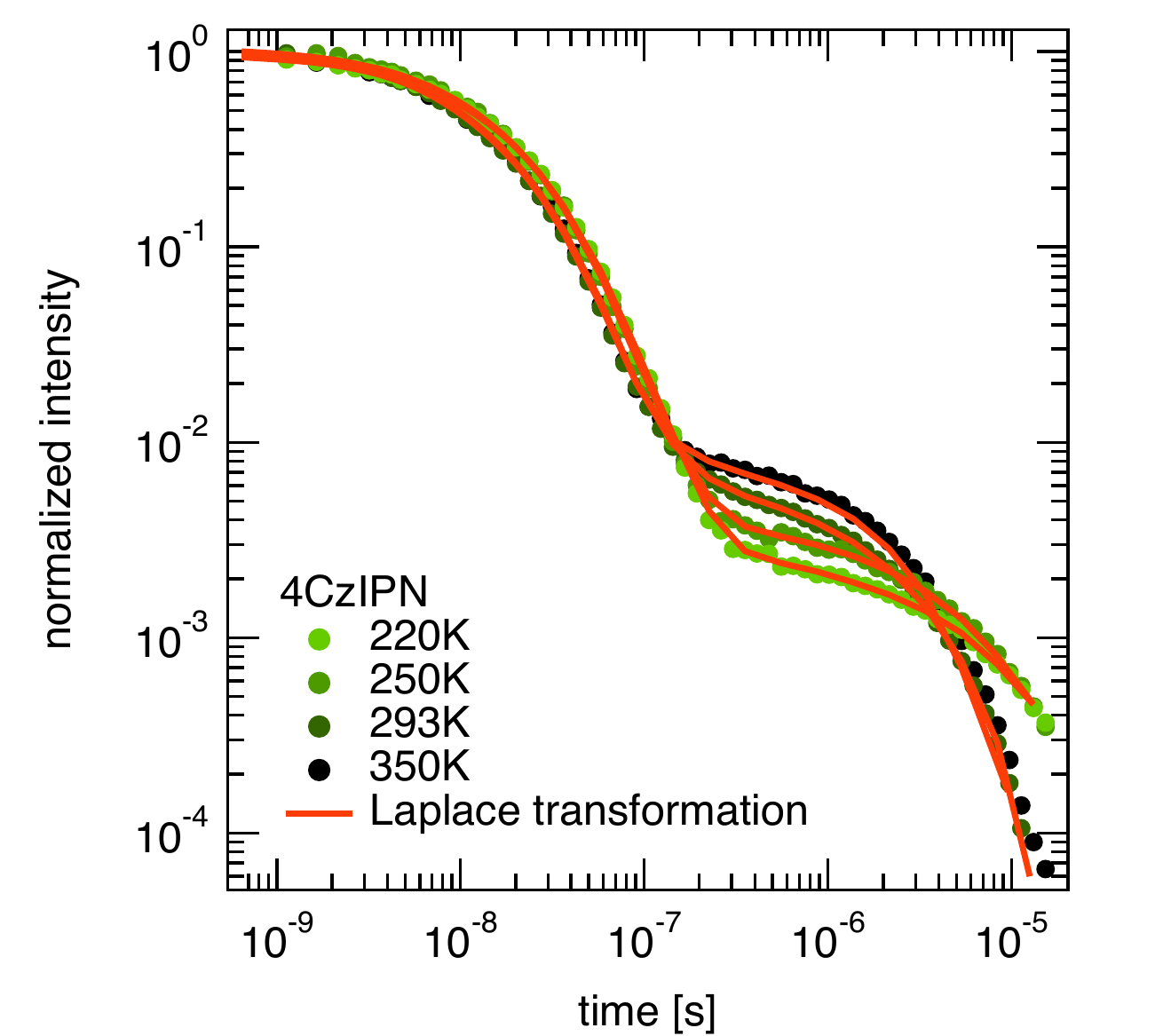}}
\qquad
\subfigure[$\mathrm{4CzIPN}$ in PS]{\includegraphics*[scale=0.45]{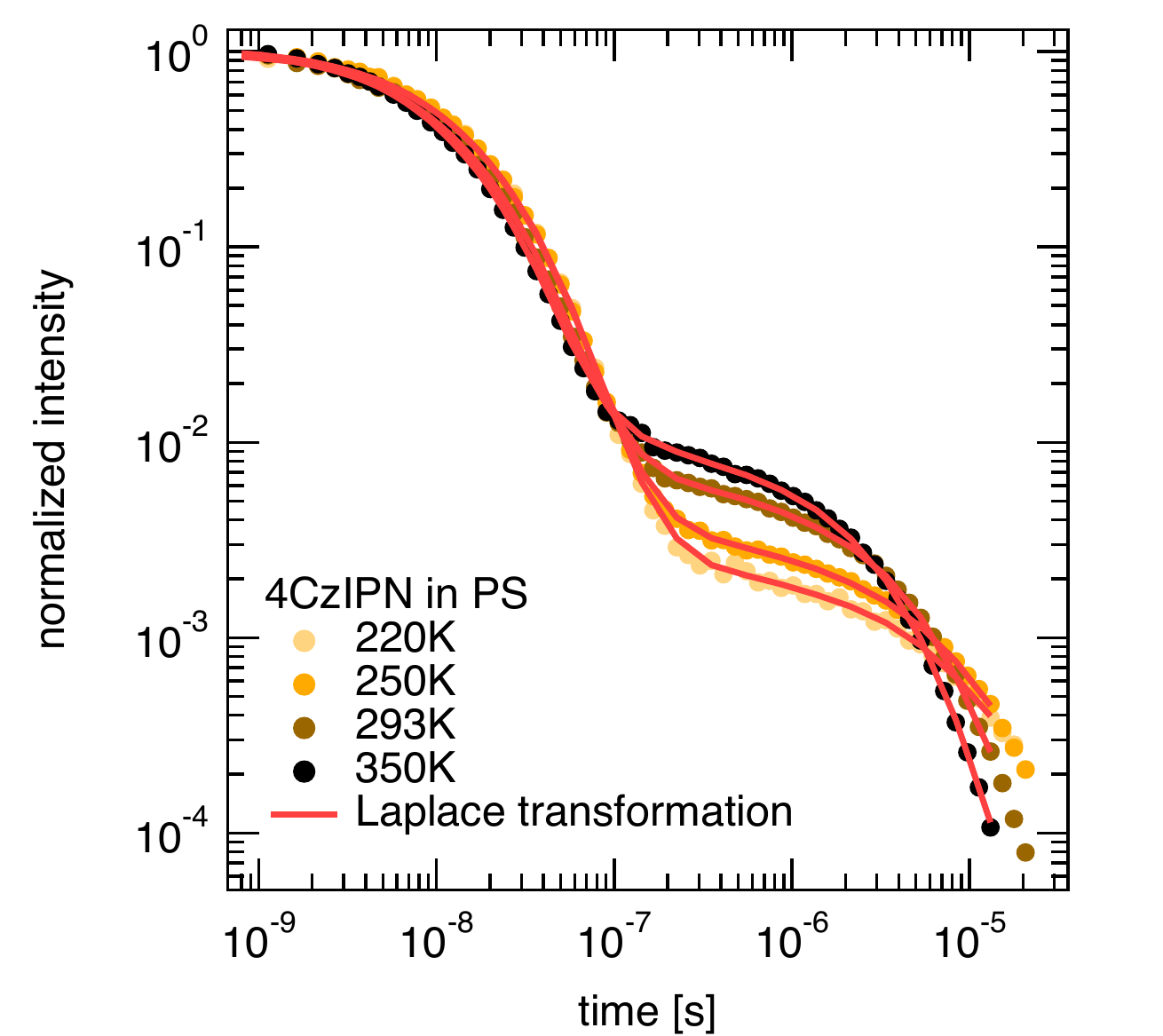}}

\centering
\subfigure[$\mathrm{3CzClIPN}$]{\includegraphics*[scale=0.45]{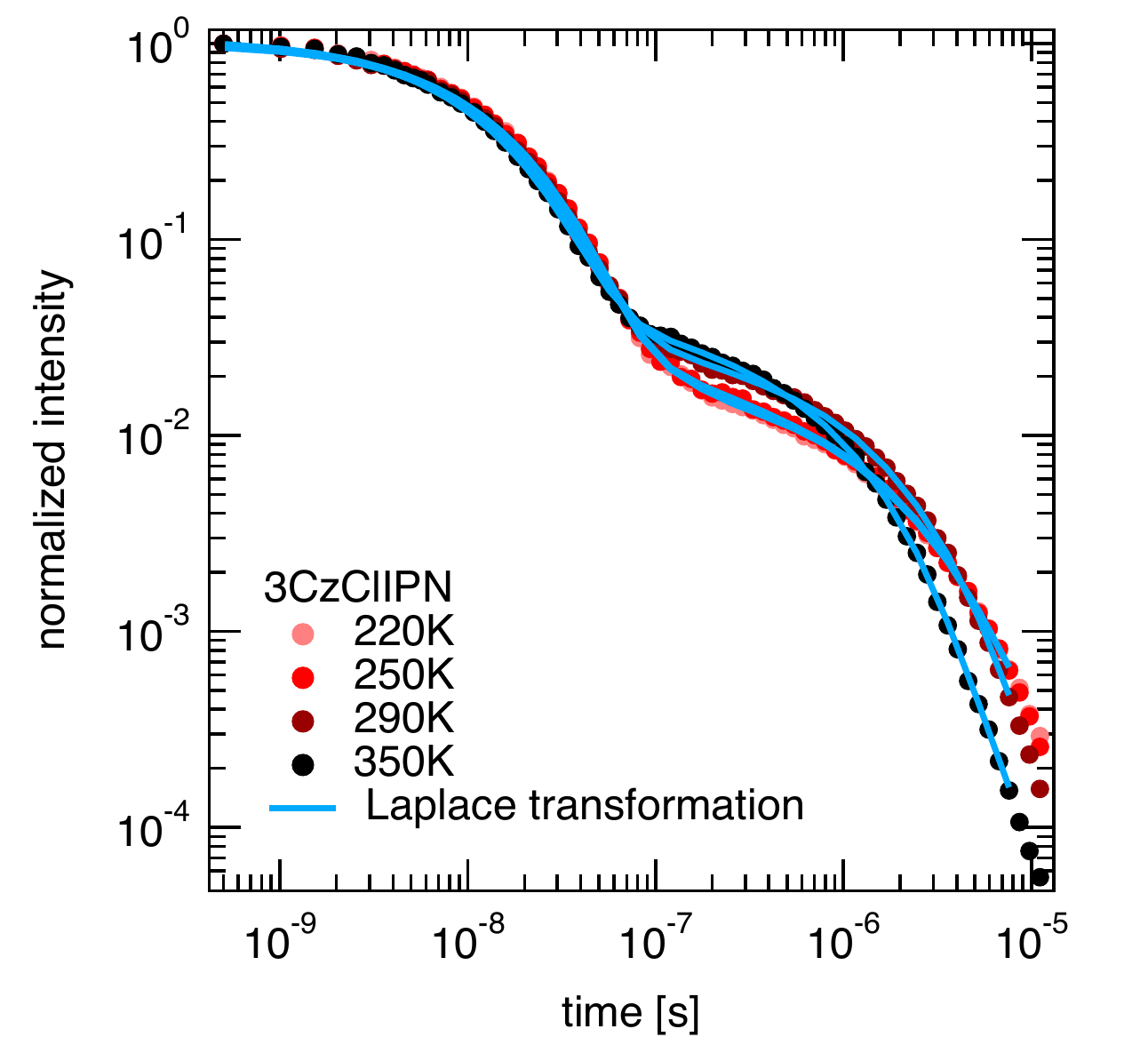}}
\qquad
\subfigure[$\mathrm{3CzClIPN}$ in PS]{\includegraphics*[scale=0.45]{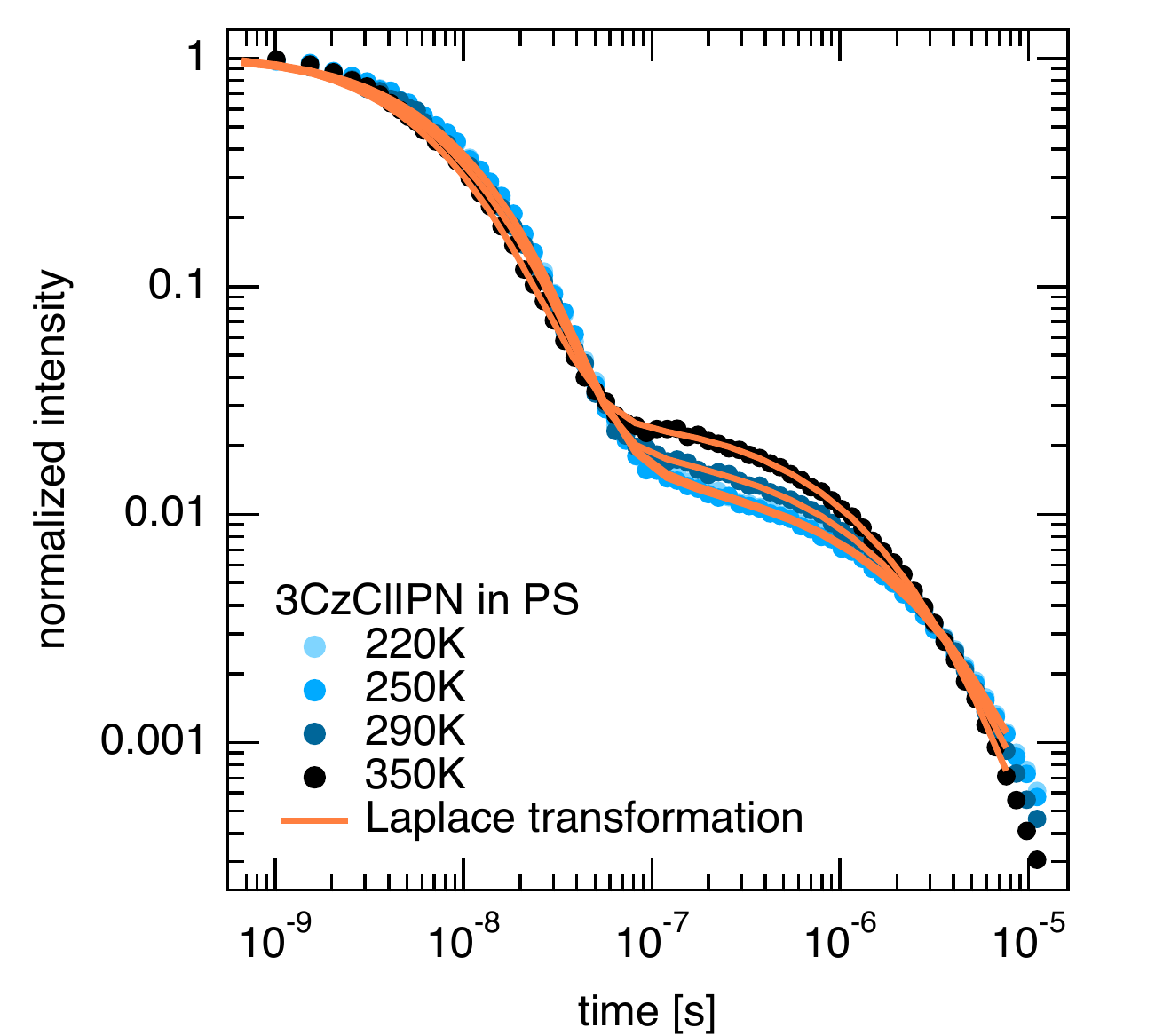}}
\caption{Laplace transform fits}
\label{laptranfit}
\end{figure}

\clearpage

\begin{figure}[h]
\centering
\subfigure[$\mathrm{4CzIPN}$]{\includegraphics*[scale=0.45]{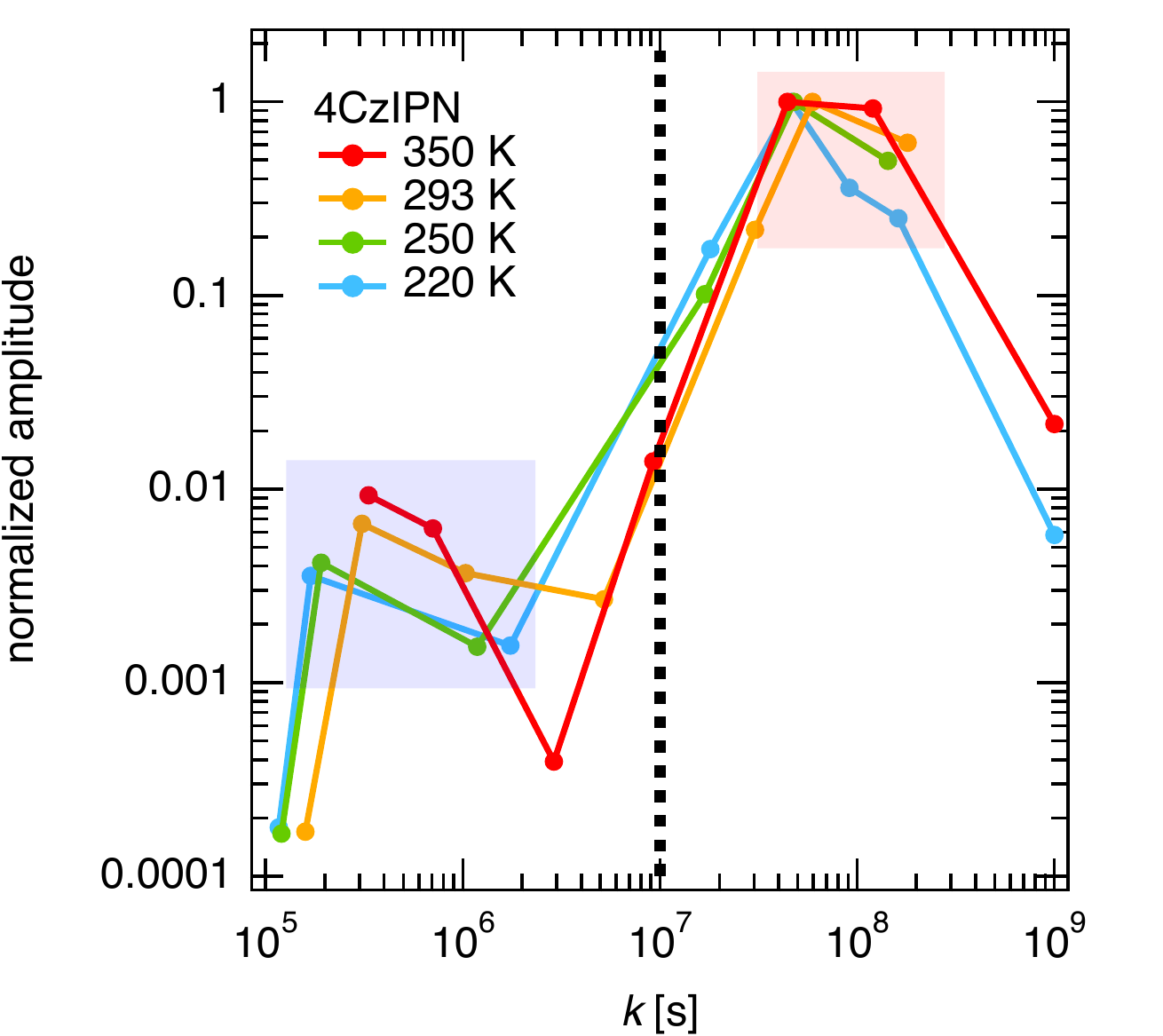}}
\qquad
\subfigure[$\mathrm{4CzIPN}$ in PS]{\includegraphics*[scale=0.45]{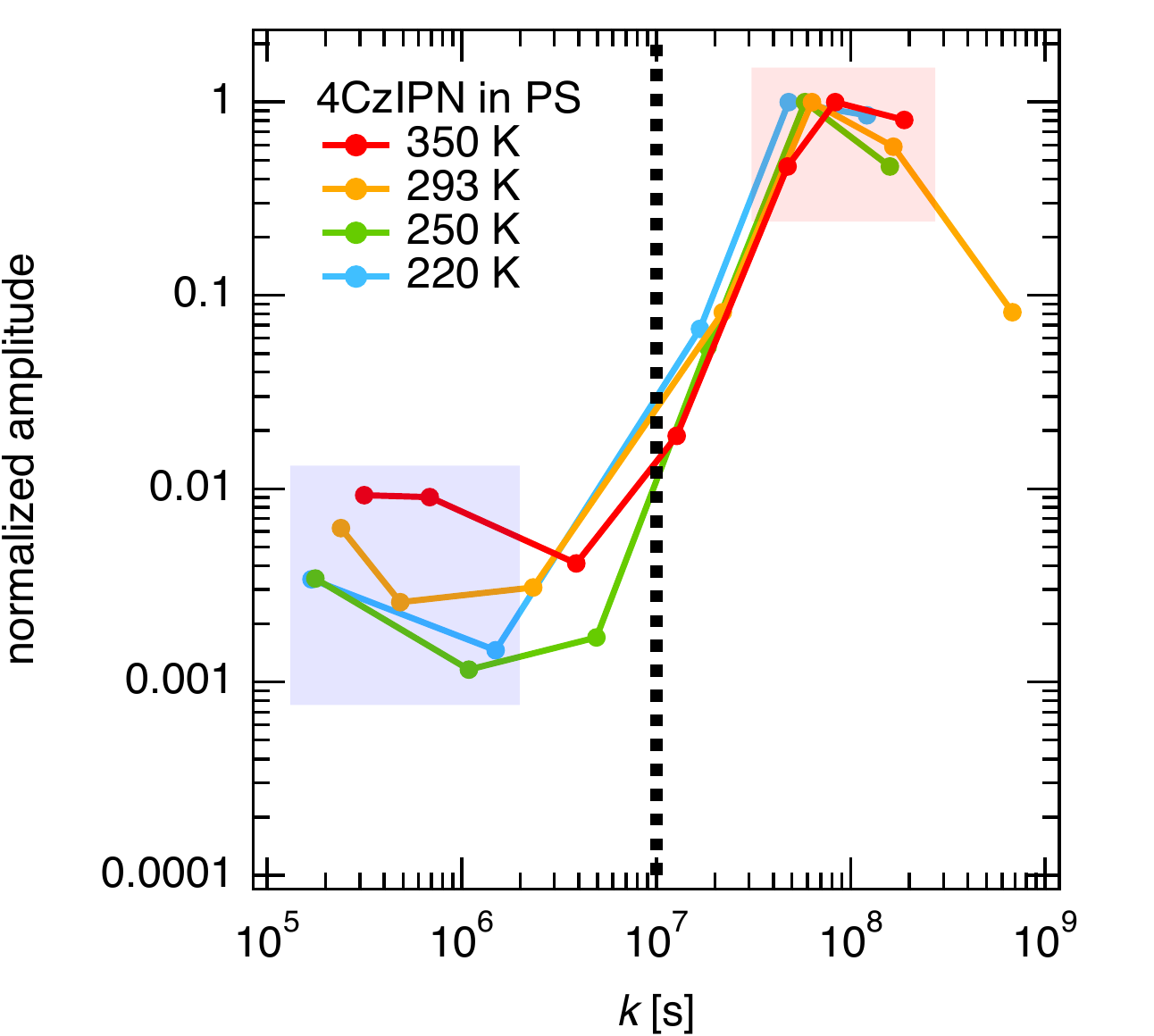}}

\centering
\subfigure[$\mathrm{3CzClIPN}$]{\includegraphics*[scale=0.45]{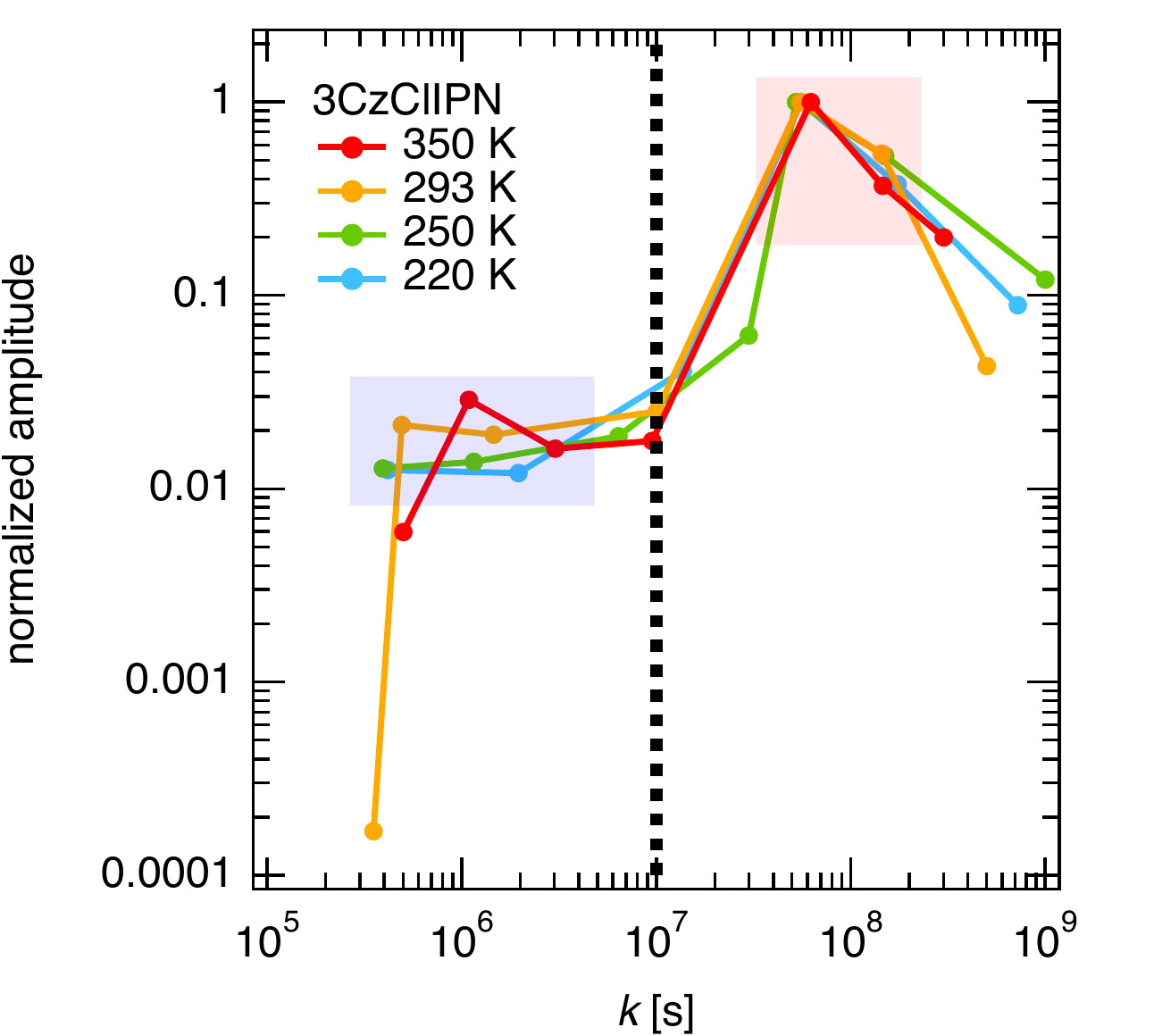}}
\qquad
\subfigure[$\mathrm{3CzClIPN}$ in PS]{\includegraphics*[scale=0.45]{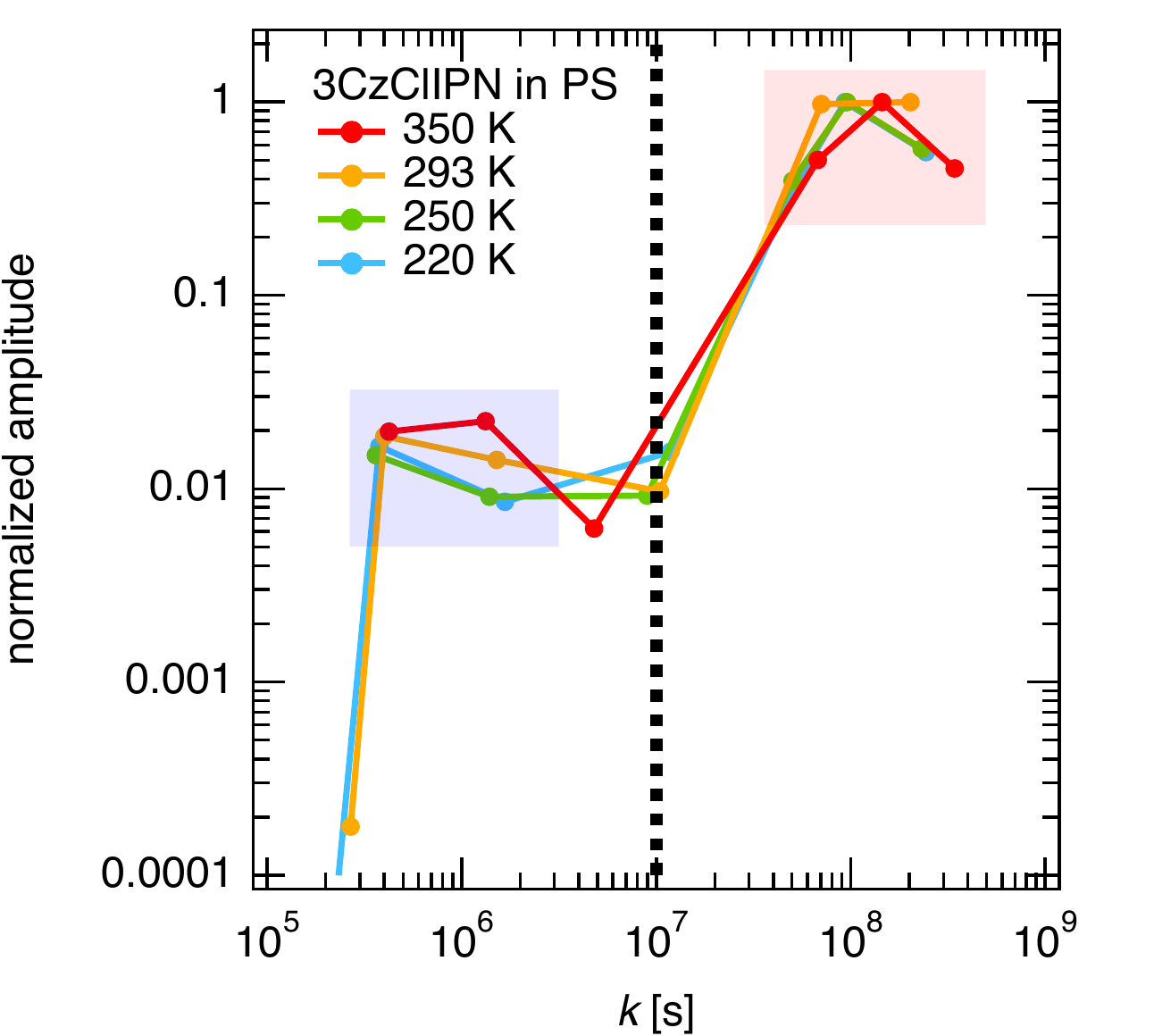}}
\caption{Laplace rates and amplitudes}
\label{laprates}
\end{figure}

\begin{figure}[h]
\centering
\subfigure[]{\includegraphics*[scale=0.44]{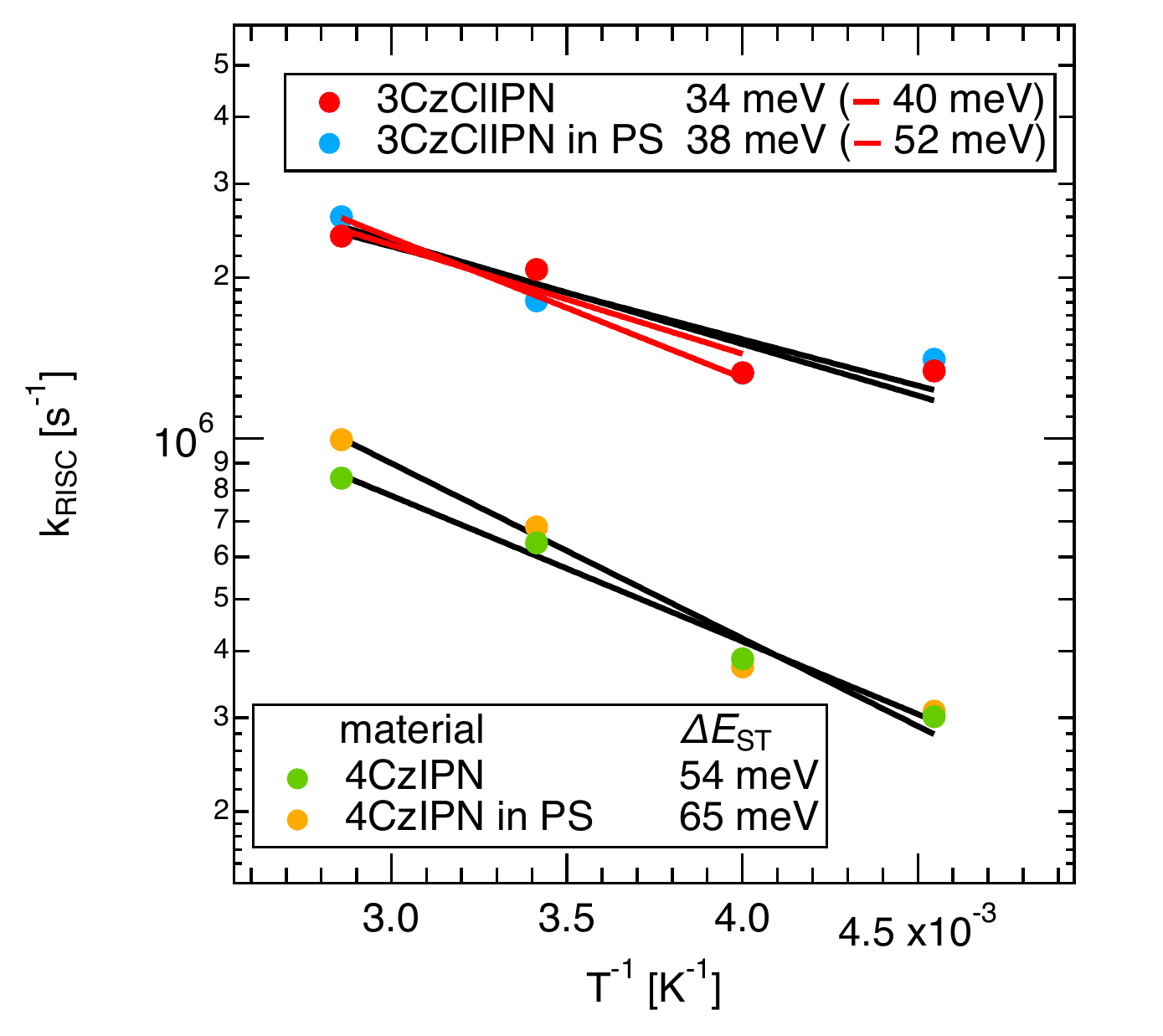}}
\qquad
\subfigure[]{\includegraphics*[scale=0.4]{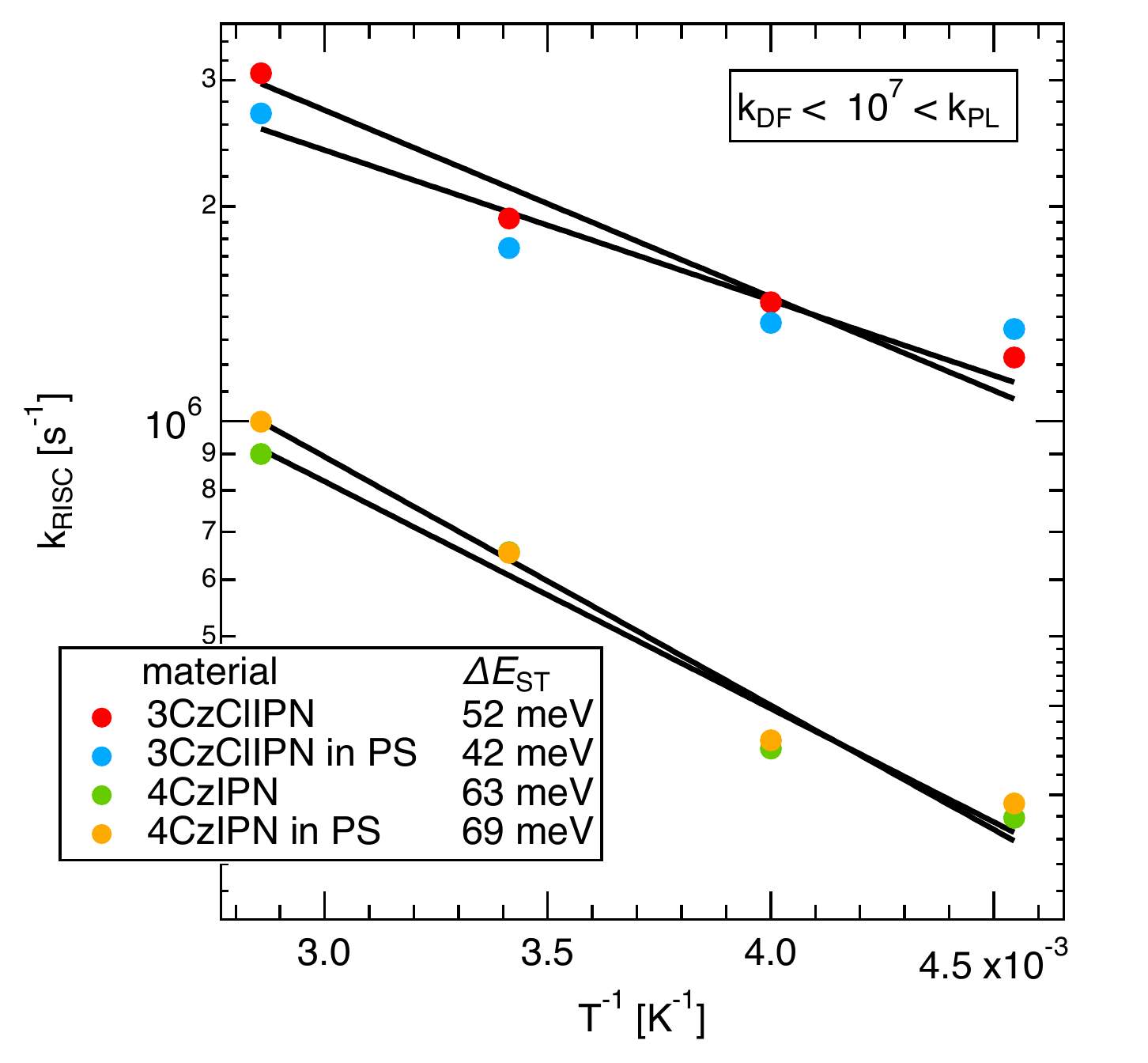}}

\caption{Arrhenius plots of Laplace analysis}
\label{arrlaplac}
\end{figure}

\clearpage 
 \section{Alternative Rate differential equation fits}
 
\noindent It is possible to achieve precise fits which fit the stretched exponential part of the delayed fluorescence by  summing up two singlet states $S_{1}(t) + S_{2}(t)$. Both $S(t)$ functions are the solution of the follwing coupled differential equation system with  individual depopulation rates $k_{\mathrm{F}}$, reverse intersystem crossing rates $k_{\mathrm{RISC}}$ and one shared intersystem crossing rate $k_{\mathrm{ISC}}$:

\begin{align}
    \dot S_{1}(t) &= - \left(k_{\mathrm{F1}}+k_{\mathrm{ISC}}\right) S_{1}(t) + k_{\mathrm{RISC1}}T_{1}(t)\\
    \dot T_{1}(t) &= k_{\mathrm{ISC}}S_{1}(t) - k_{\mathrm{RISC1}}T_{1}(t)\\
    \dot S_{2}(t) &= - \left(k_{\mathrm{F2}}+k_{\mathrm{ISC}}\right) S_{2}(t) + k_{\mathrm{RISC2}}T_{2}(t)\\
    \dot T_{2}(t) &= k_{\mathrm{ISC}}S_{2}(t) - k_{\mathrm{RISC2}}T_{2}(t).
\label{globalfitequ}
\end{align}

\subsection{Fits}
 \begin{figure}[h]
 \centering
 \subfigure[$\mathrm{4CzIPN}$]{\includegraphics*[scale=0.5]{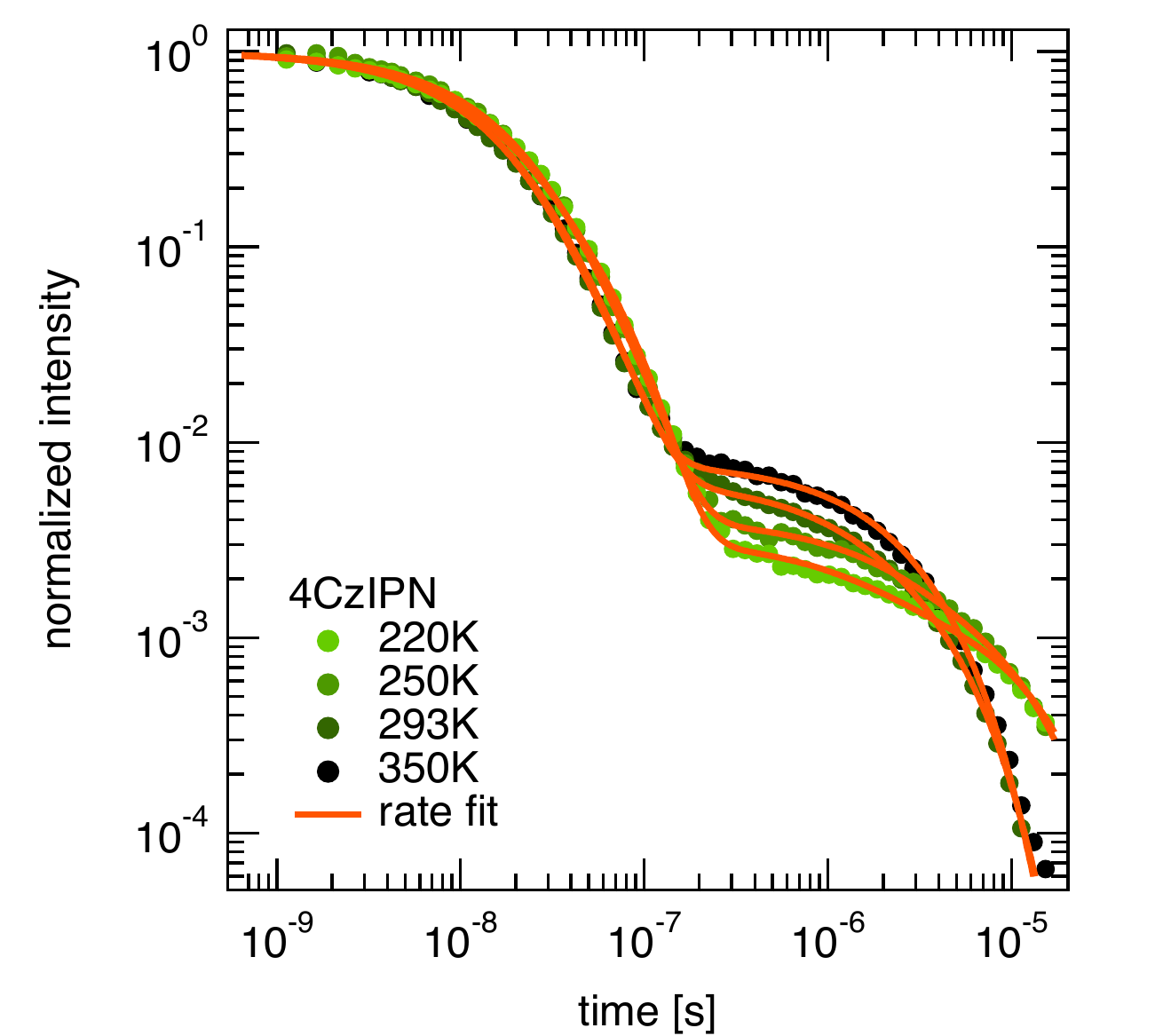}}
 \qquad
 \subfigure[$\mathrm{4CzIPN}$ in PS]{\includegraphics*[scale=0.5]{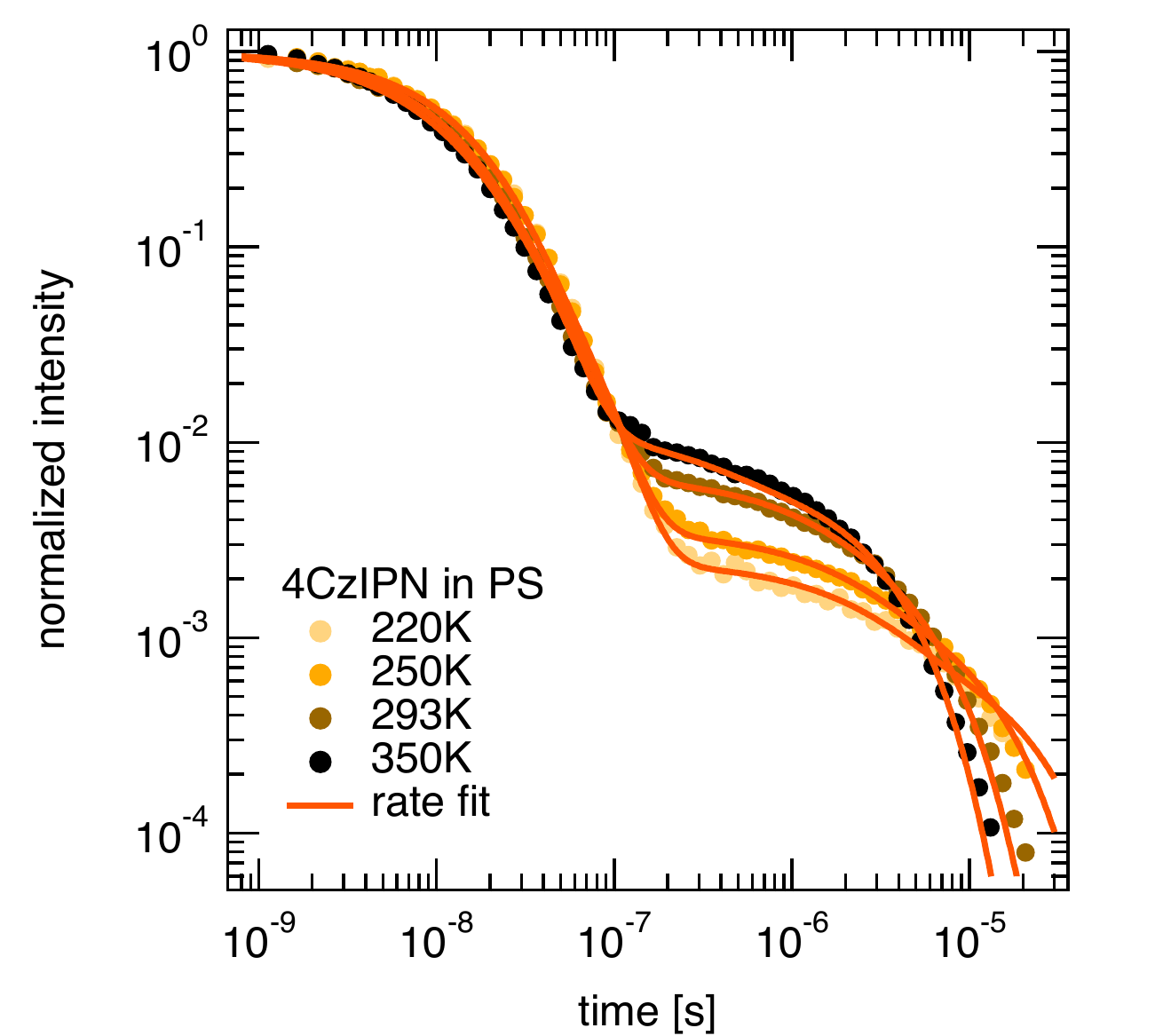}}

 \subfigure[$\mathrm{3CzClIPN}$]{\includegraphics*[scale=0.5]{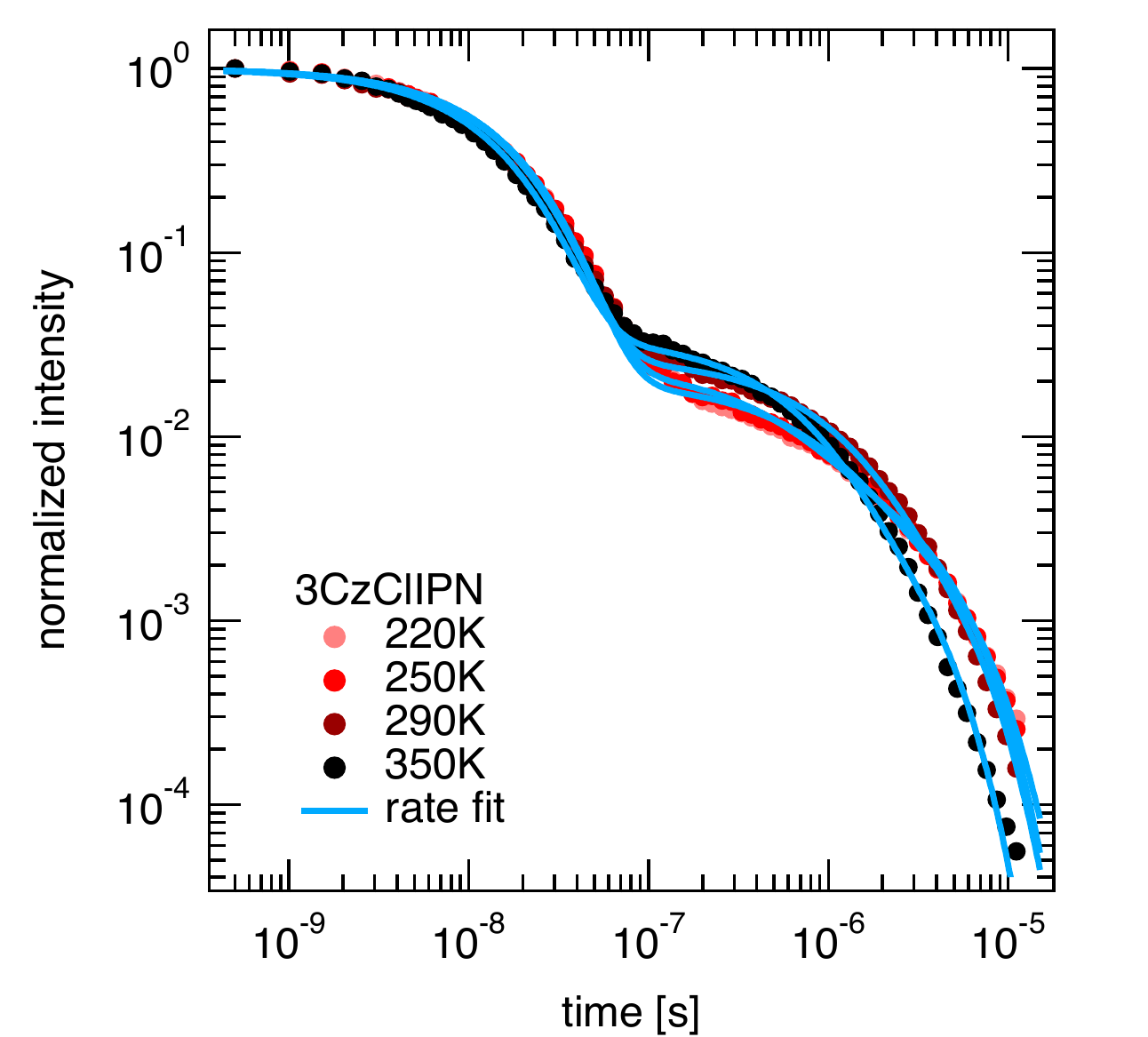}}
 \qquad
 \subfigure[$\mathrm{3CzClIPN}$ in PS]{\includegraphics*[scale=0.5]{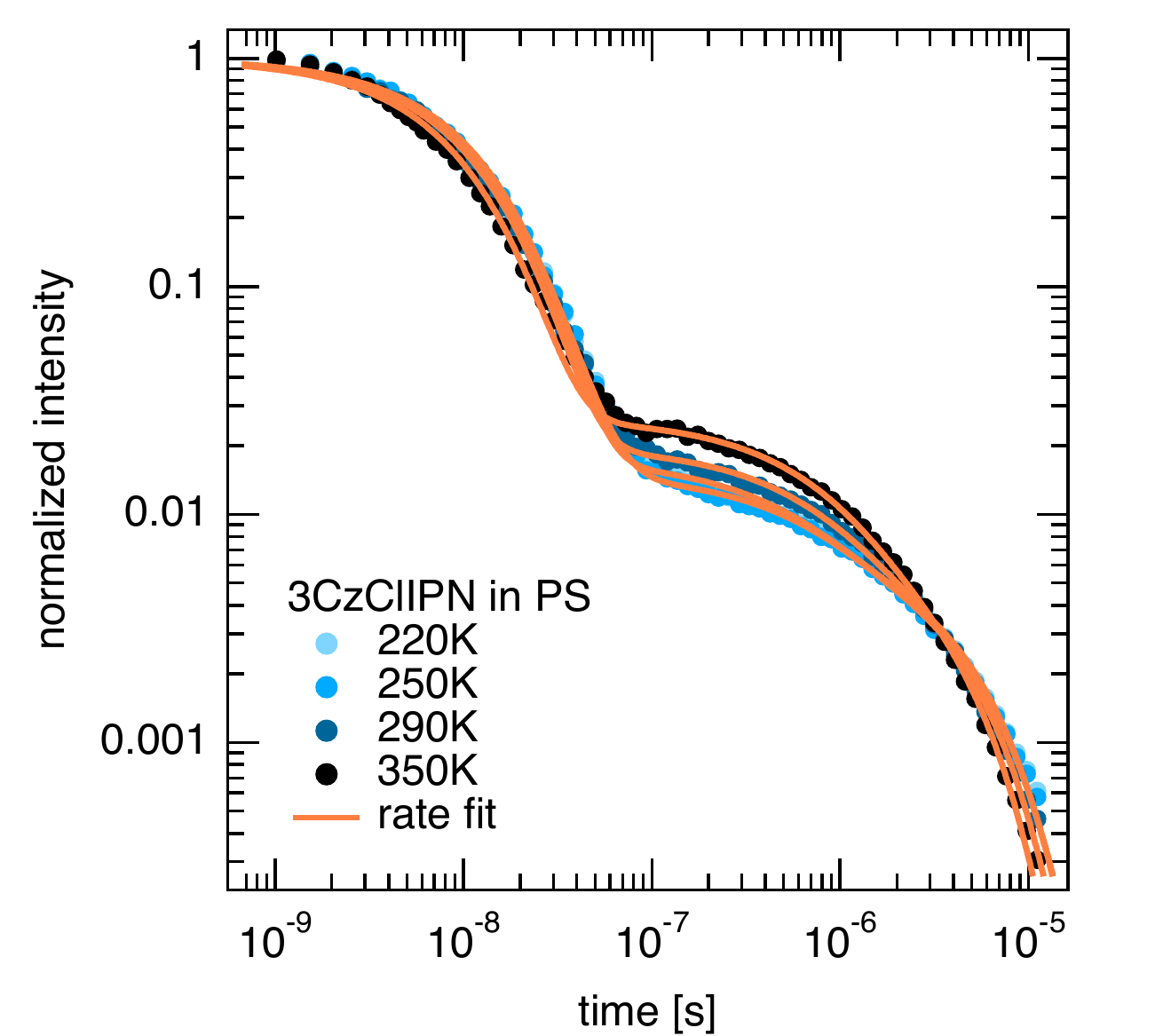}}
 \caption{Distribution of $k_{\mathrm{RISC}}$}
 \label{distr}
 \end{figure}

\clearpage
\subsection{Rate analysis}

\noindent Assuming two different $\Delta E_{\mathrm{ST}}$ was suggested by Kobayashi \textit{et al.} for neat $\mathrm{4CzIPN}$ films.\cite{kobayashi2017contributions} However, it is not clear if interpreting the Arrenhius plot is useful for two different $k_{\mathrm{RISC}}$ (or the corresponding weighted Arithmetic or geometric means as shown in figures B a-d) because they are not connected. Therefore, the deviation between both $k_{\mathrm{RISC}}$ looks differently for each sample, although all fits have a similar quality. This also makes treating the approach as a global fit difficult.  Interpreting the temperature-dependent  $k^1_{\mathrm{RISC}}$ and $k^2_{\mathrm{RISC}}$ individually leads to values which are not plausible. The geometric mean deviates from the values presented in the main paper without a clear trend. All Arrenhius fits with $k_{\mathrm{RISC}} = k_{\mathrm{A}} \exp{(-\Delta E_{\mathrm{ST}} / k_{\mathrm{B}} T)} $ are summarized in table \ref{arrh}.

 \begin{figure}[h]
 \centering
 \subfigure[$\mathrm{4CzIPN}$]{\includegraphics*[scale=0.5]{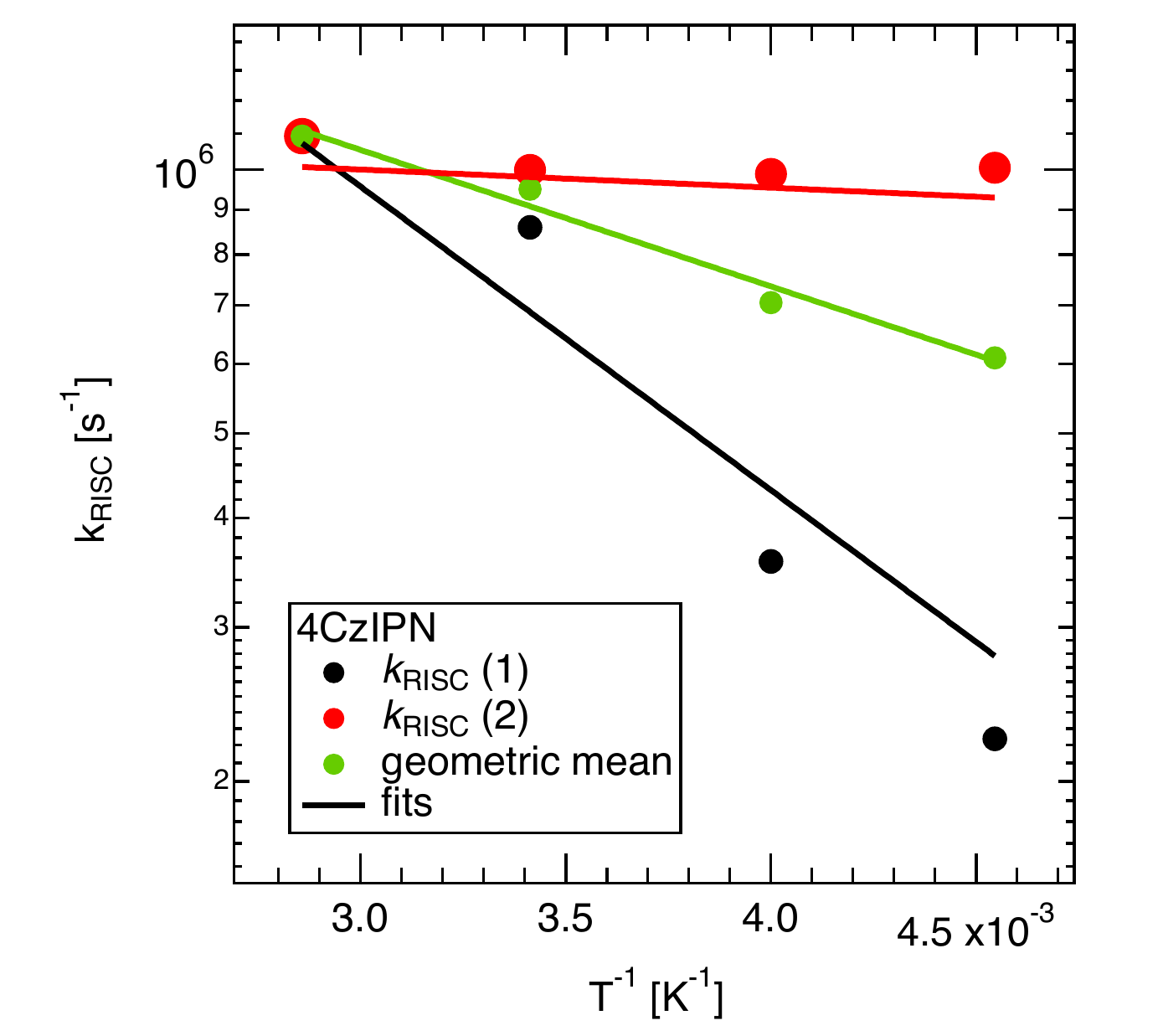}}
 \qquad
 \subfigure[$\mathrm{4CzIPN}$ in PS]{\includegraphics*[scale=0.5]{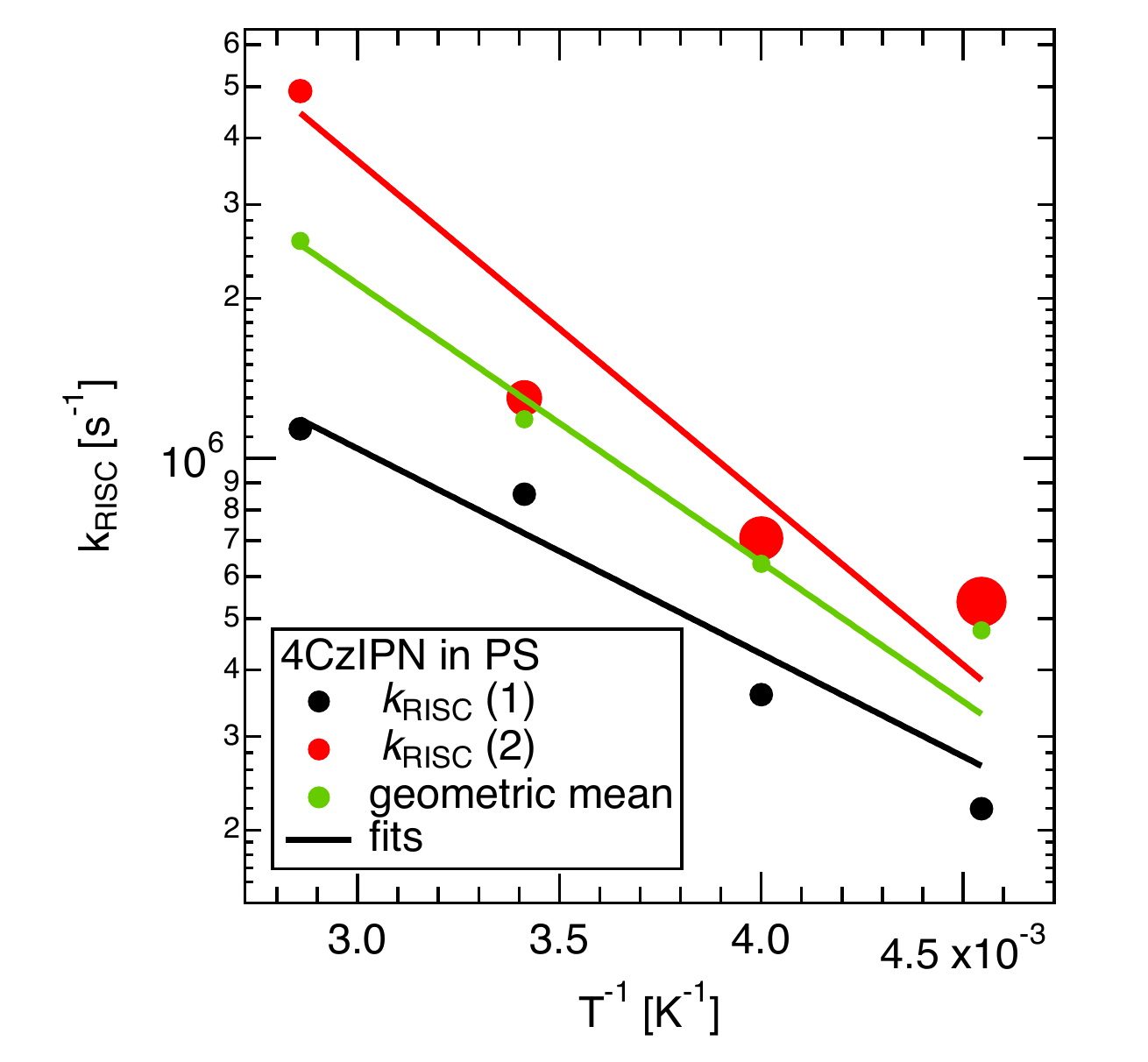}}

 \subfigure[$\mathrm{3CzClIPN}$]{\includegraphics*[scale=0.5]{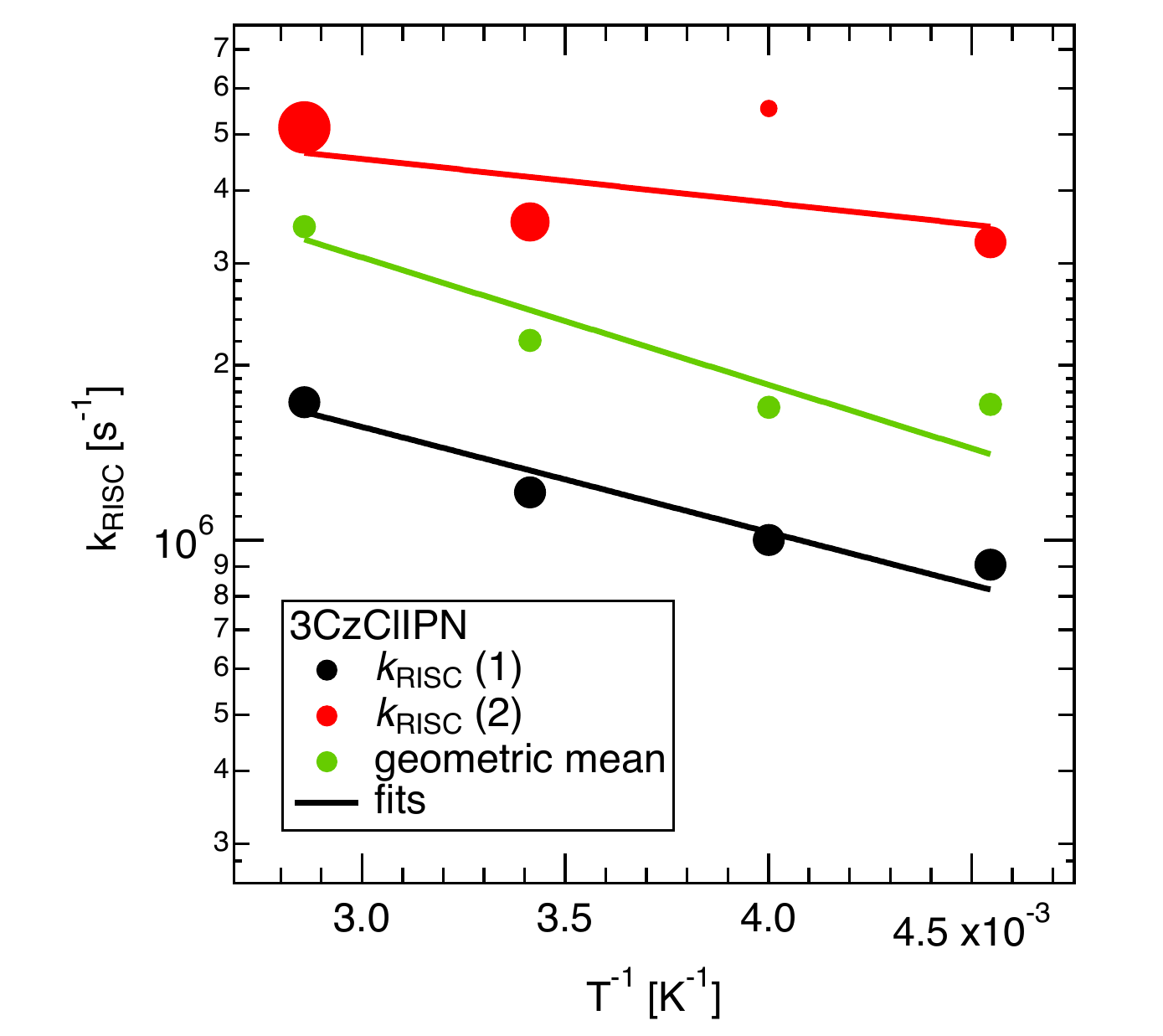}}
 \qquad
 \subfigure[$\mathrm{3CzClIPN}$ in PS]{\includegraphics*[scale=0.5]{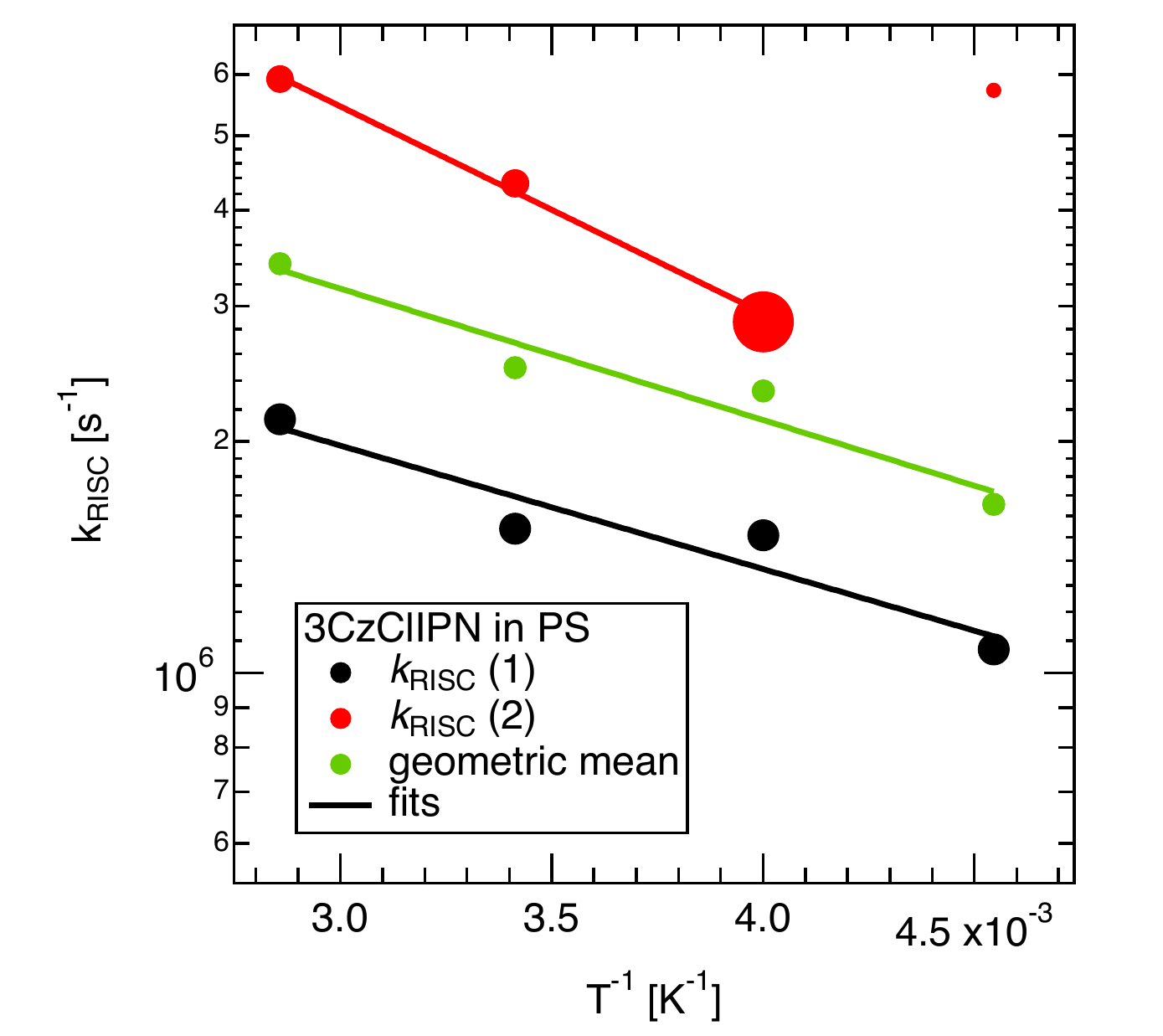}}
 \caption{Weighted (dot size) $k_{\mathrm{RISC}}$ rates from rate fits.}
 \label{weight2r}
 \end{figure}

\begin{table} [h]
\begin{tabular}{llcccc} 

 && $\mathrm{4CzIPN}$  &   & $\mathrm{3CzClIPN}$  &   \\
 && neat film &  in PS &  neat film &  in PS \\
 \hline
 $\Delta E_{\mathrm{ST}}(1)$&$[\mathrm{meV}]$ & 69 &76 &36&32\\
 $\Delta E_{\mathrm{ST}}(2)$&$[\mathrm{meV}]$ & 4 & 125&15&53\\
 $\Delta E_{\mathrm{ST}}(\mathrm{geo})$&$[\mathrm{meV}]$ &31 &104 &43&34\\
 
\end{tabular}

\caption{Individual $\Delta E_{\mathrm{ST}}$ and weighted geometric means of the Arrenhius plots.}
\label{arrh}
\end{table}

\clearpage 
\section{Global fits}

\begin{figure}[h]
\centering
\subfigure[$\mathrm{4CzIPN}$]{\includegraphics*[scale=0.45]{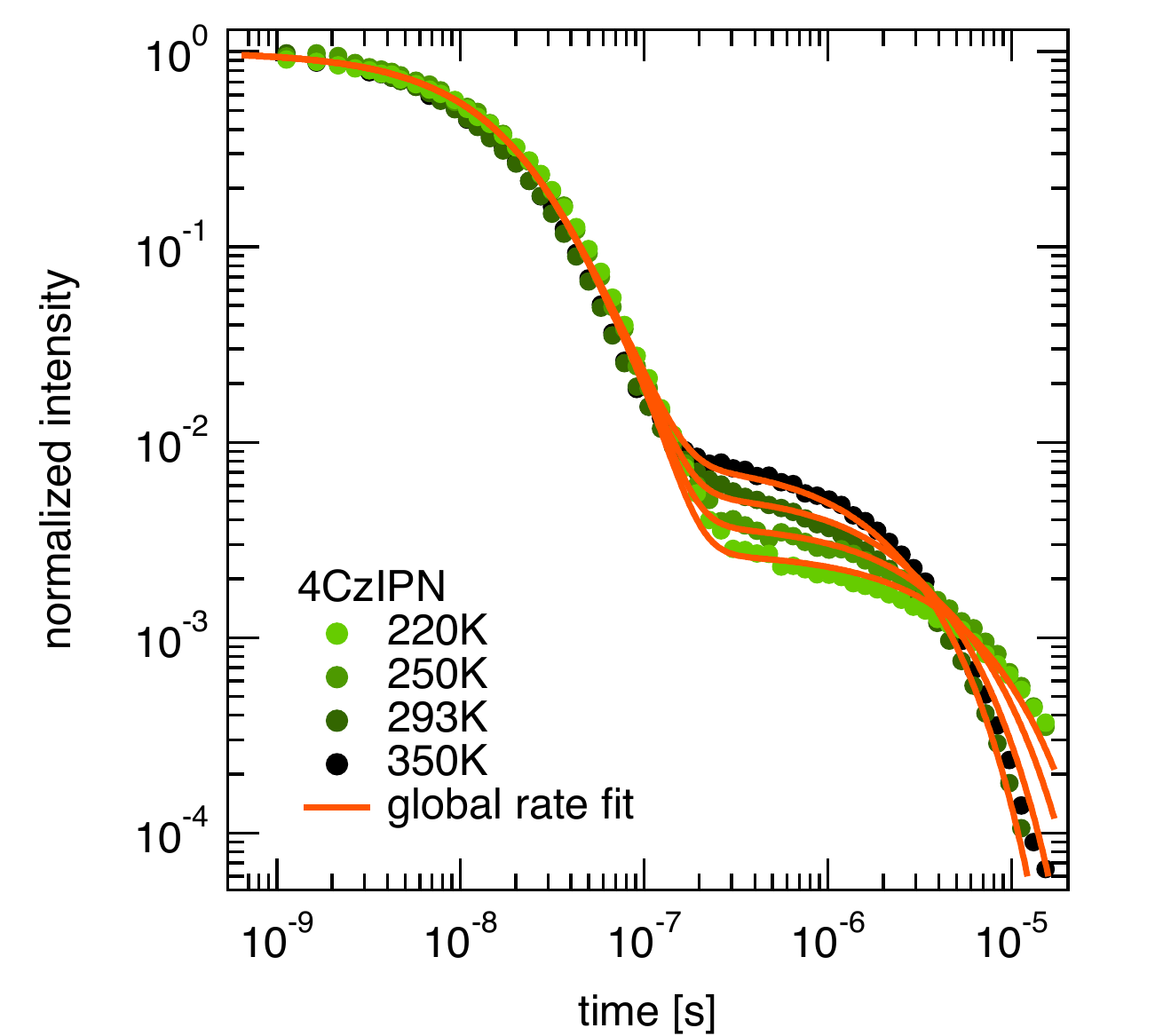}}
\qquad
\subfigure[$\mathrm{4CzIPN}$ in PS]{\includegraphics*[scale=0.45]{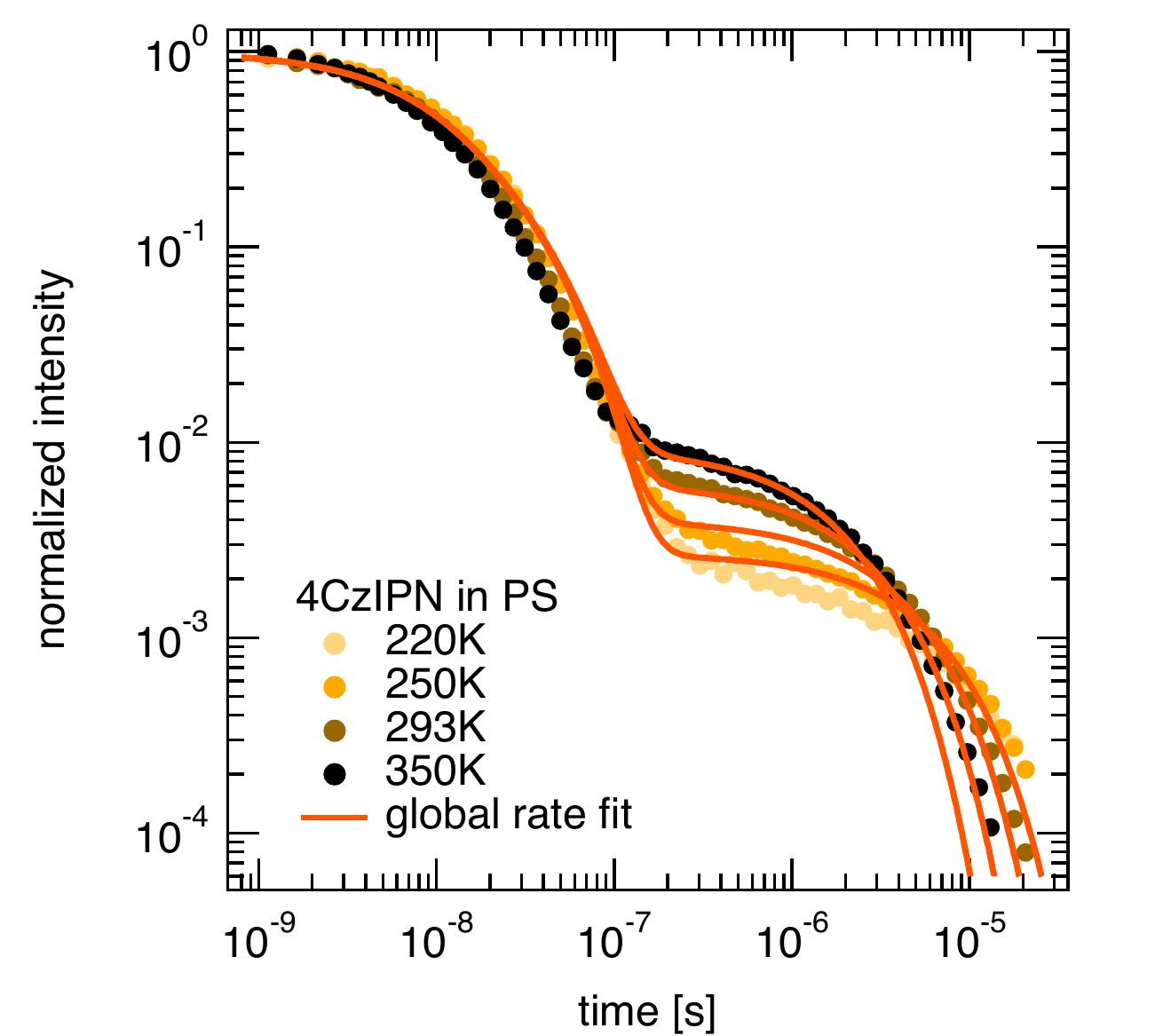}}

\centering
\subfigure[$\mathrm{3CzClIPN}$]{\includegraphics*[scale=0.45]{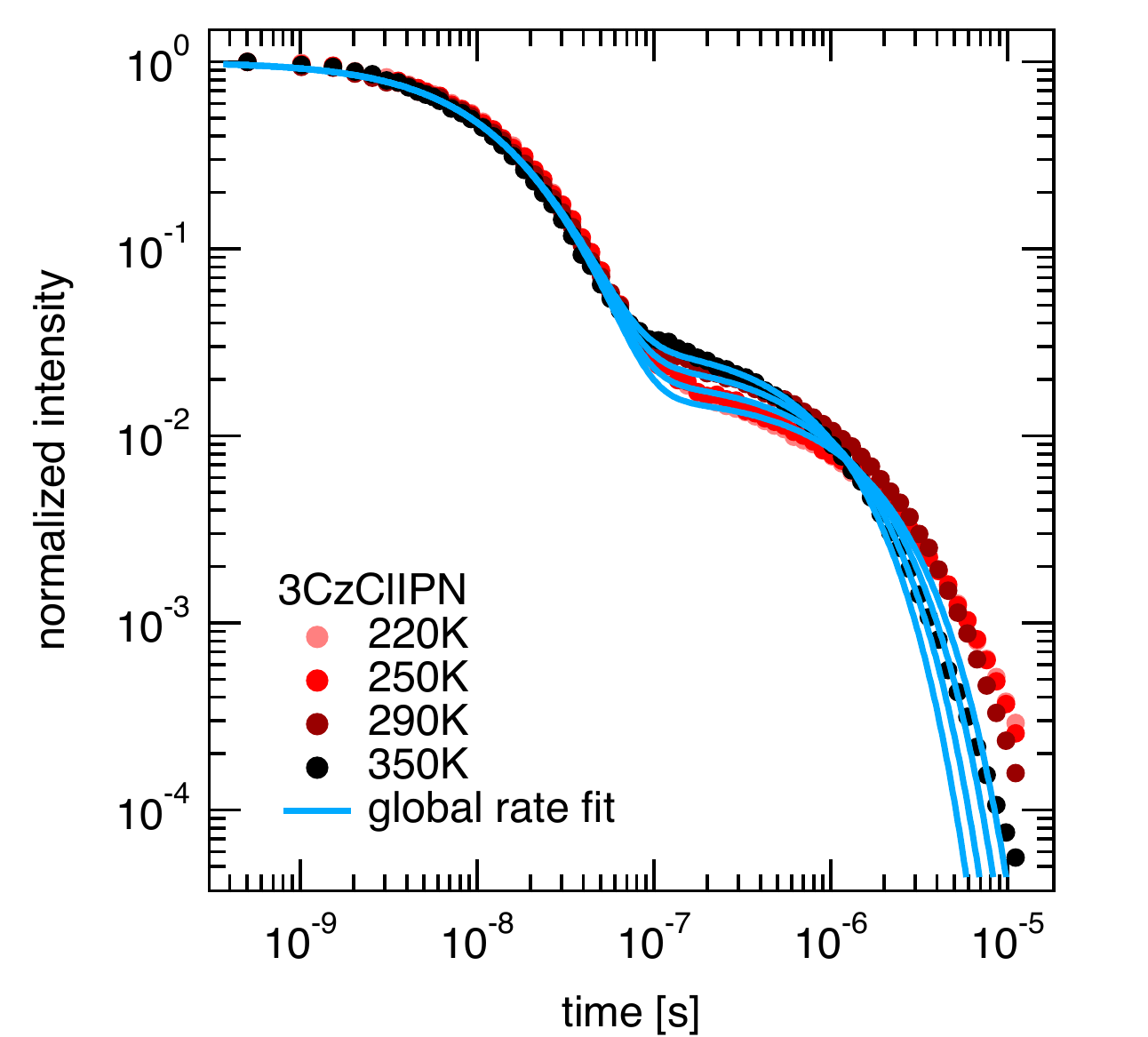}}
\qquad
\subfigure[$\mathrm{3CzClIPN}$ in PS]{\includegraphics*[scale=0.45]{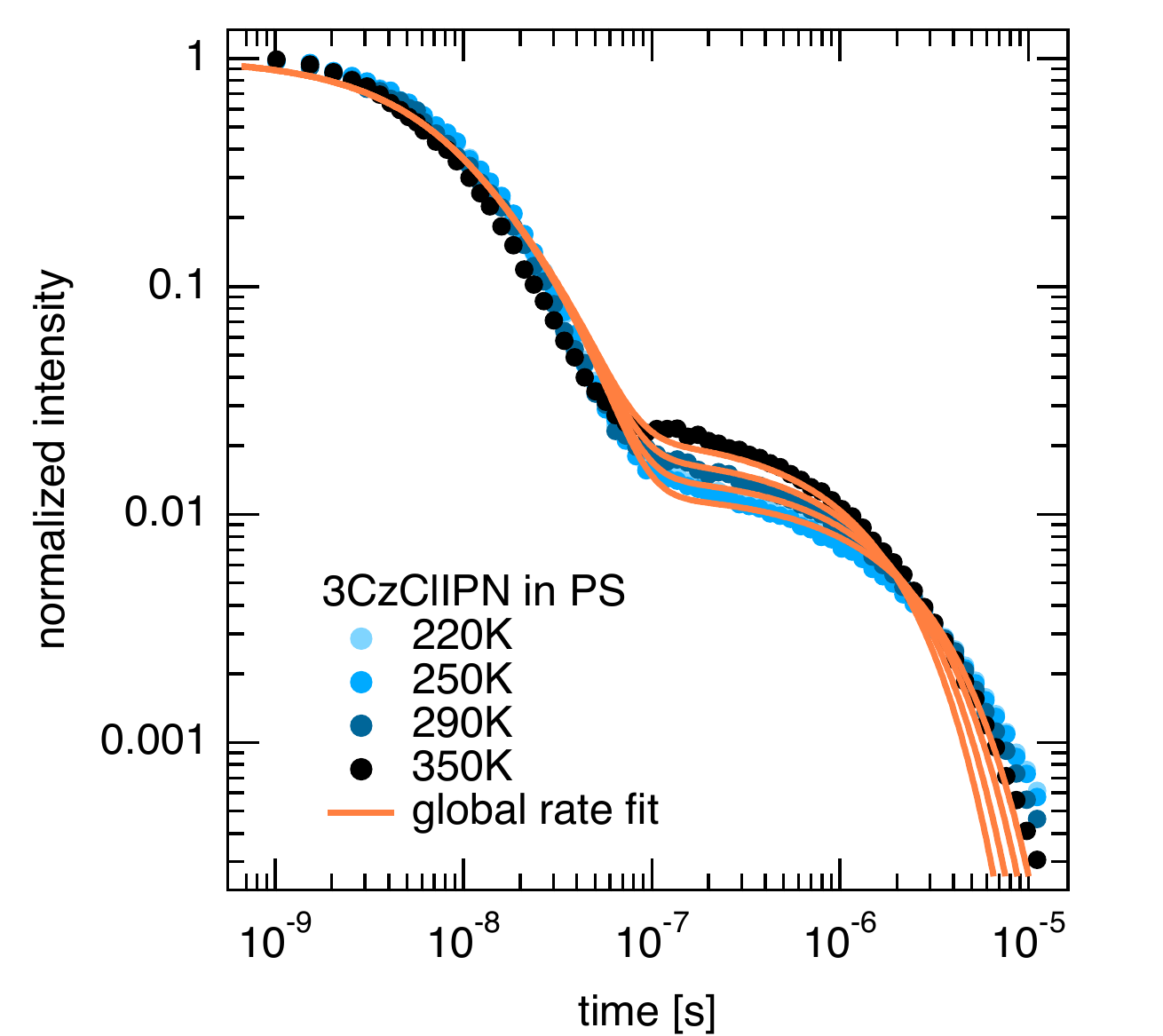}}
\caption{global fits}
\label{globfitsfig}
\end{figure}

  \begin{figure}[ht]
  \centering
\includegraphics*[scale=0.4]{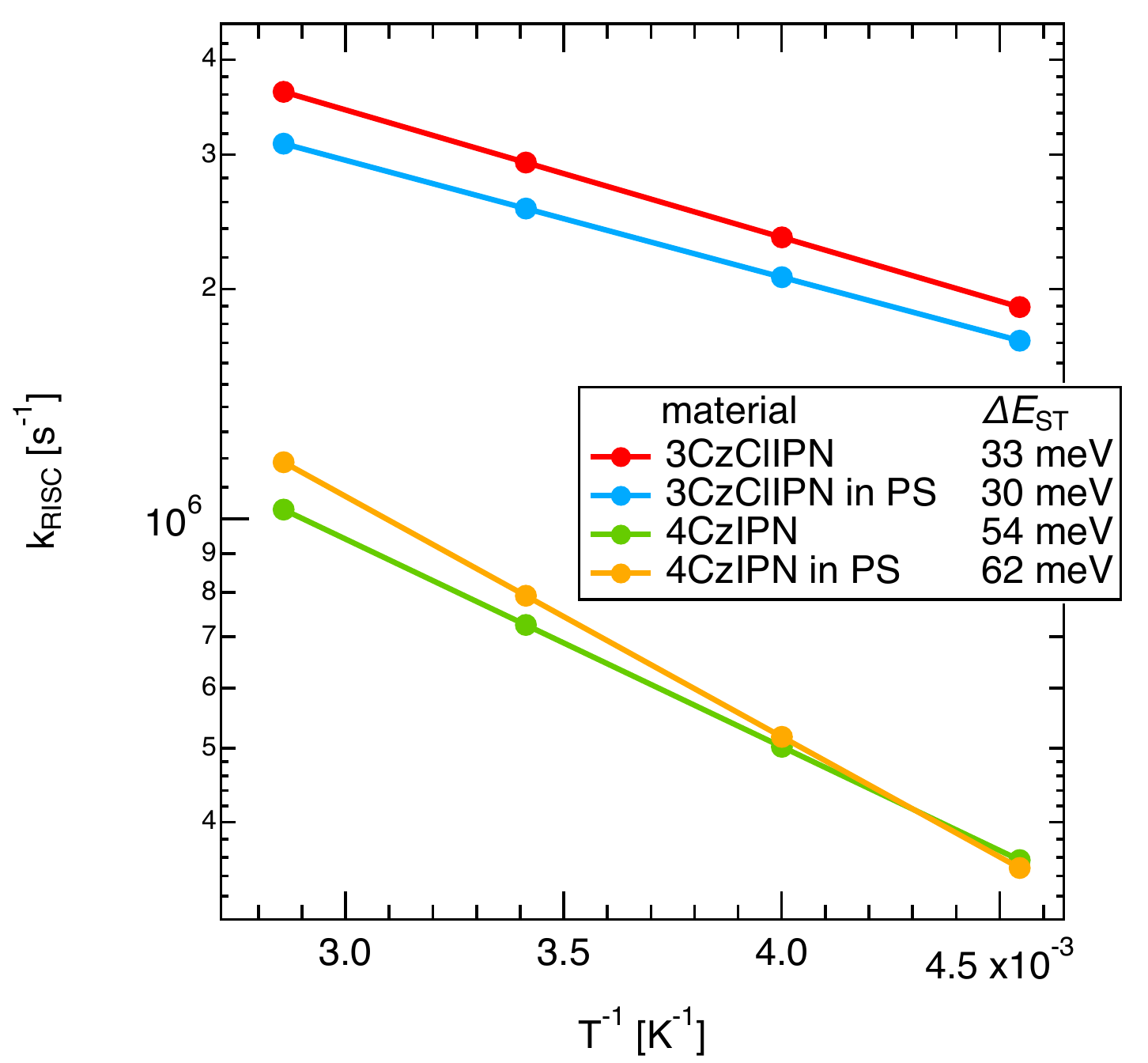}
  \caption{Arrhenius plot of global fits}
    \label{arrglob}
    \end{figure}

\clearpage
\section{Steady-state measurements}

\begin{figure}[h]
\centering
\subfigure[Comparison of $5\,\mathrm{wt\%}$ $\mathrm{4CzIPN}$ and $\mathrm{3CzClIPN}$ in CBP dissolved in toluene. CBP concentration $1\,\mathrm{mg\,ml^{-1}}$.]{\includegraphics*[scale=0.5]{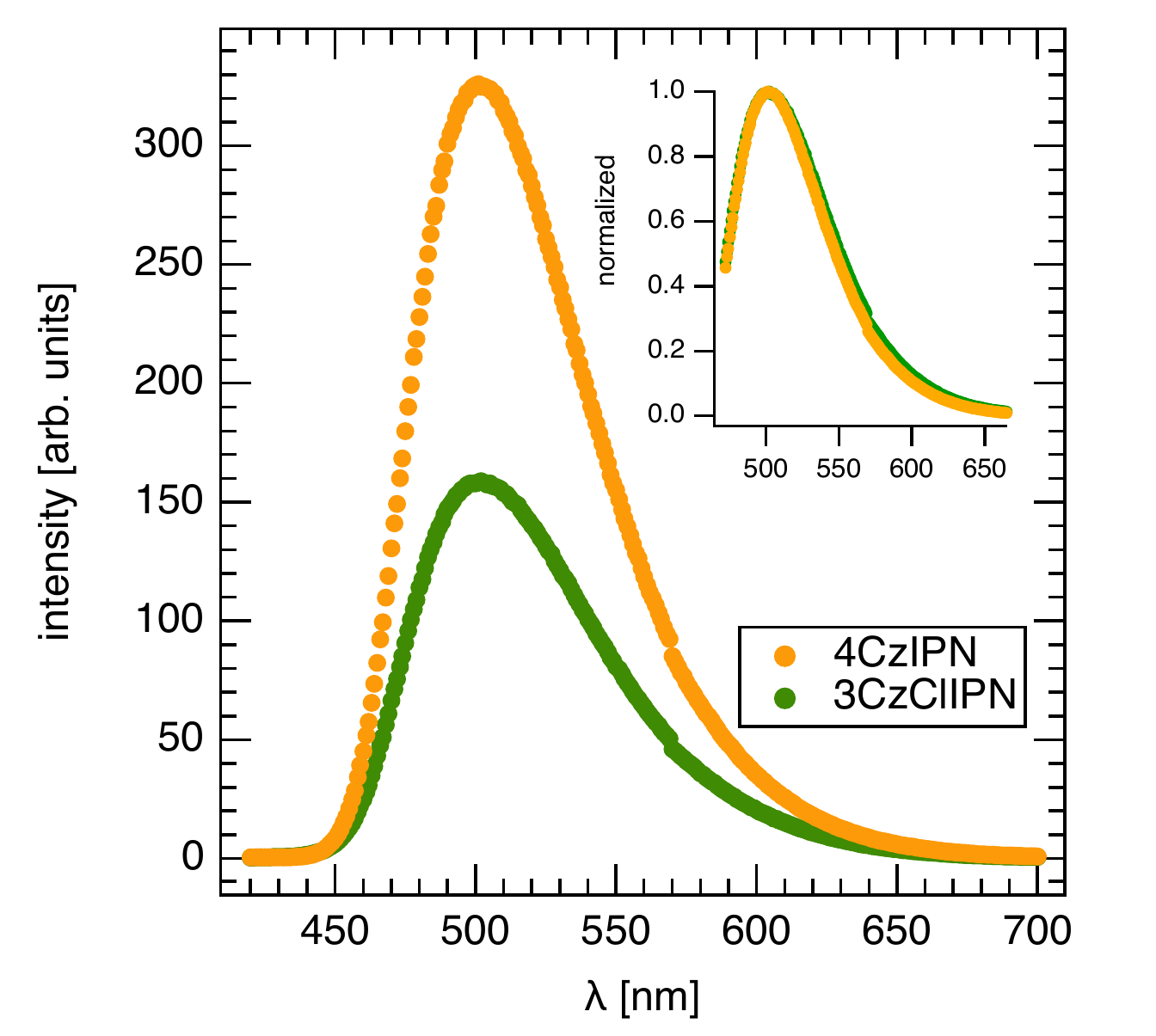}}
\qquad
\subfigure[Normalized steady-state PL spectra of neat and doped PS films on glass indicating red-shift for neat material due to reabsorption.]{\includegraphics*[scale=0.5]{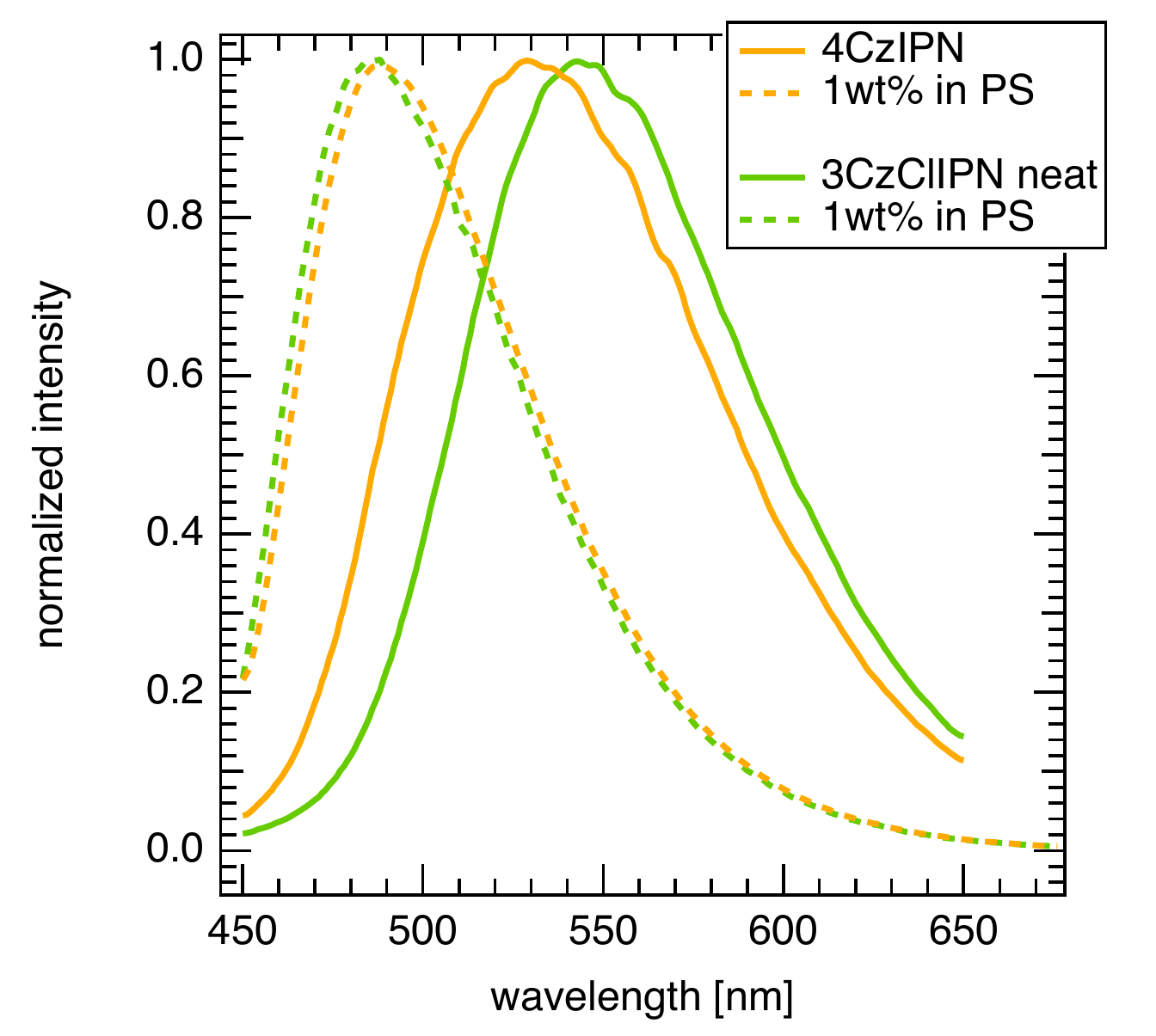}}

\subfigure[Photoluminescence quantum yield (PLQY: $\mathrm{4CzIPN}$ 99\,\%, $\mathrm{3CzClIPN}$ 54\,\%)]{\includegraphics*[scale=0.5]{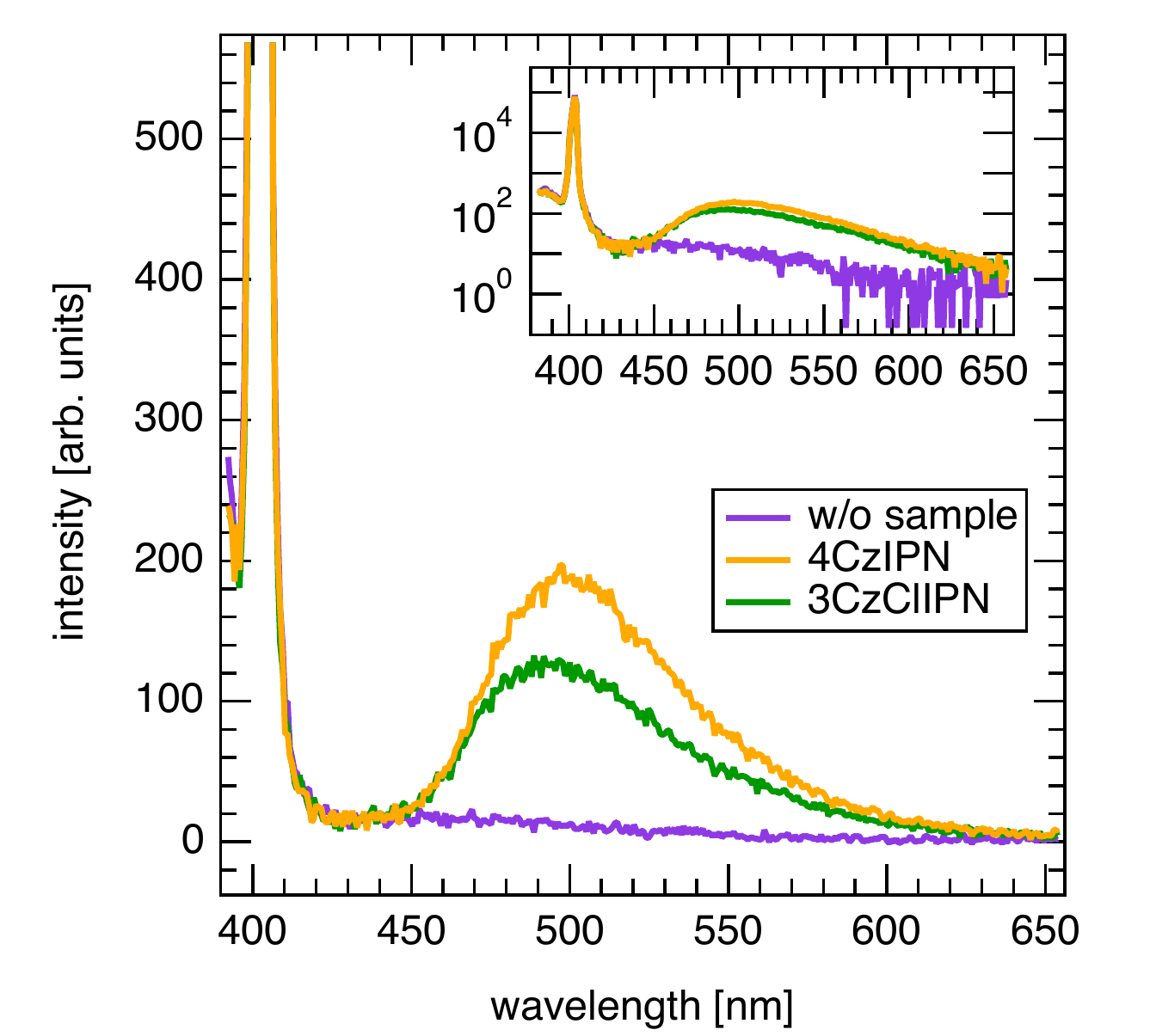}}
\qquad
\subfigure[Normalized electroluminescence embedded in CBP]{\includegraphics*[scale=0.5]{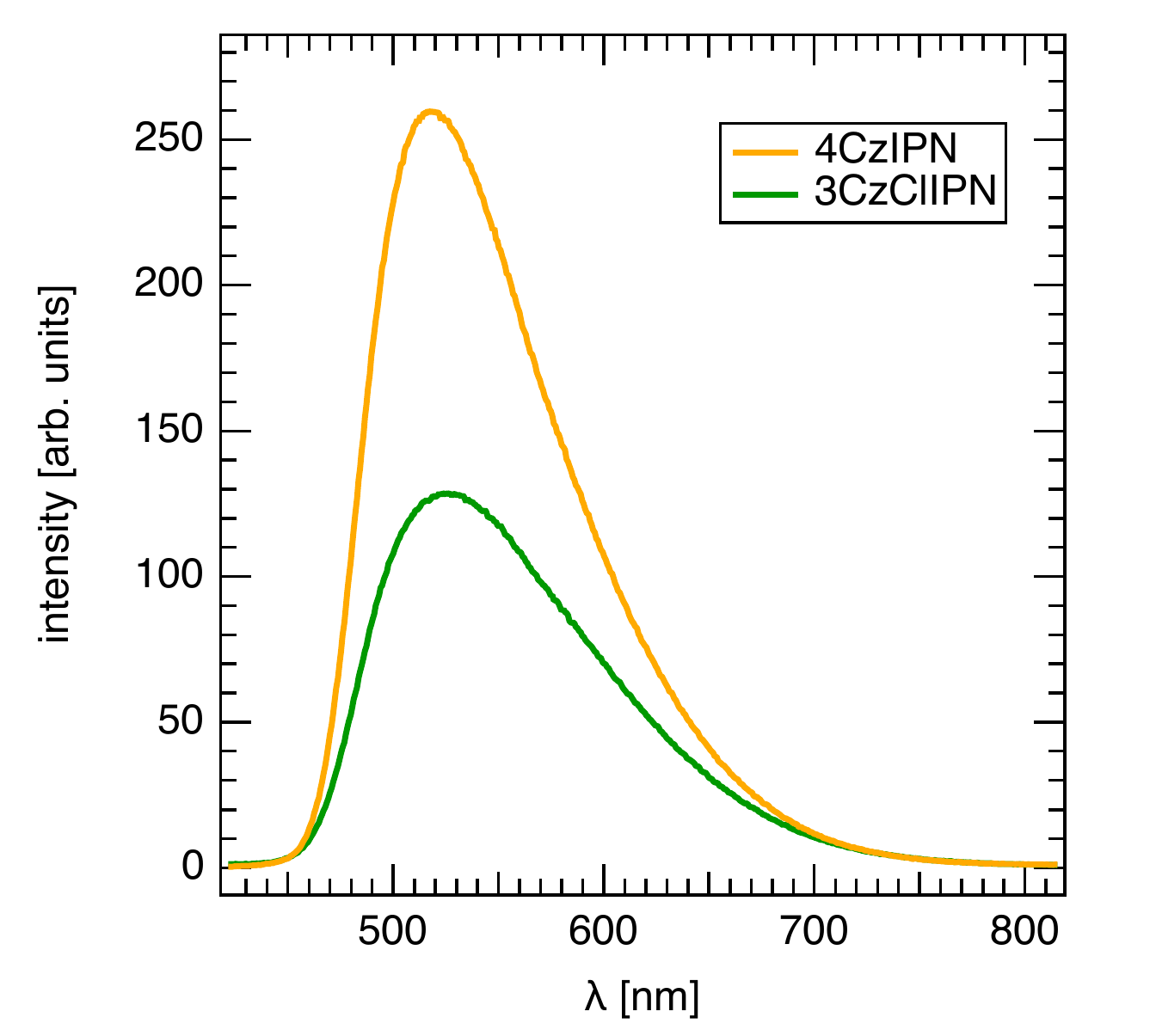}}
\caption{Steady-state spectra}
\end{figure}

\noindent Spectra of $\mathrm{4CzIPN}$ and $\mathrm{3CzClIPN}$ in solution and as films were measured with a Cary Eclipse spectrometer (Varian) and excited at 405\,nm. The absolute photoluminescence quantum yield (PLQY) was determined by measuring the spectrum (OceanOptics, QE pro - calibrated with an ORIEL 63358 tungsten lamp) of the excitation laser diode (Thorlabs, 405\,nm) and the 1\,wt\% $\mathrm{4CzIPN}$ ($\mathrm{3CzClIPN}$)  on glass film simultaneously. As a reference, a clean glass substrate was measured. PLQY is then calculated with (emitted PL)/(laser signal from clean glass - laser signal from TADF sample). For the measurement of electroluminescence, structured ITO substrates were spincoated with ZnO-NP (zinc oxide nanoparticles) dissolved in ethanol (50\,mg\,ml$\mathrm{^{-1}}$ at 2000\,rpm) and 5\,wt\% of $\mathrm{4CzIPN}$ ($\mathrm{3CzClIPN}$) in CBP (4,4'-Bis(\textit{N}-carbazolyl)-1,1'-biphenyl) dissolved in toluene (12\,mg\,ml$\mathrm{^{-1}}$ at 1500\,rpm). These films then were subsequently evaporated with 40\,nm HMTPD (\textit{N},\textit{N},\textit{N}',\textit{N}'-Tetrakis(3-methylphenyl)-3,3'-dimethylbenzidine), 10\,nm molybdenum oxide and 100\,nm aluminum. A voltage of 6\,V was applied and the spectrum was measured with a 200\,\textmu m, NA 0.22 fiber which was coupled to the substrate and the spectrometer (OceanOptics, QE pro).

\clearpage
 
\section{Time-resolved measurements}
\begin{figure}[h]
\centering
\subfigure[$\mathrm{4CzIPN}$ in PS]{\includegraphics*[scale=0.6]{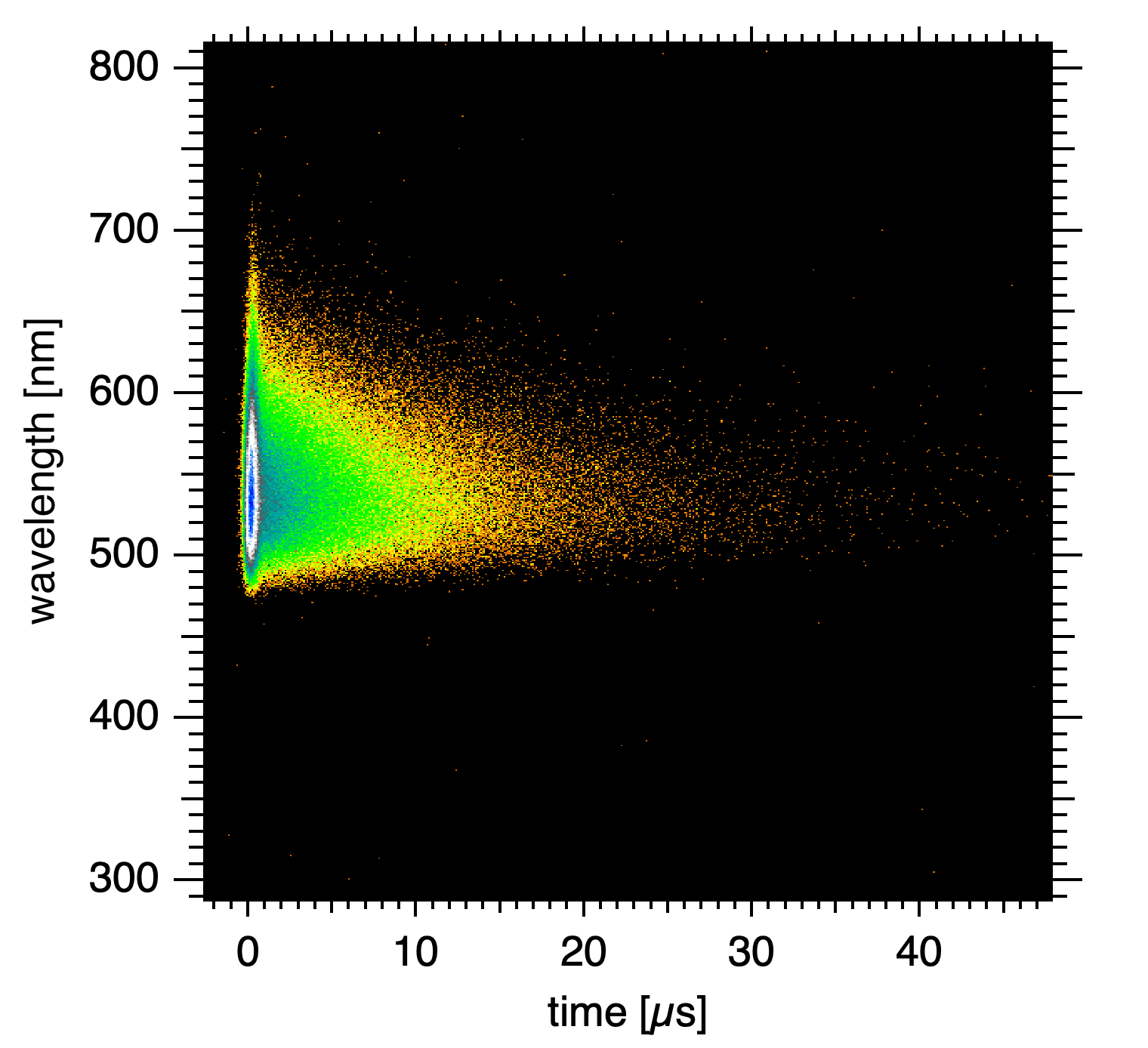}}
\qquad
\subfigure[$\mathrm{3CzClIPN}$ in PS]{\includegraphics*[scale=0.6]{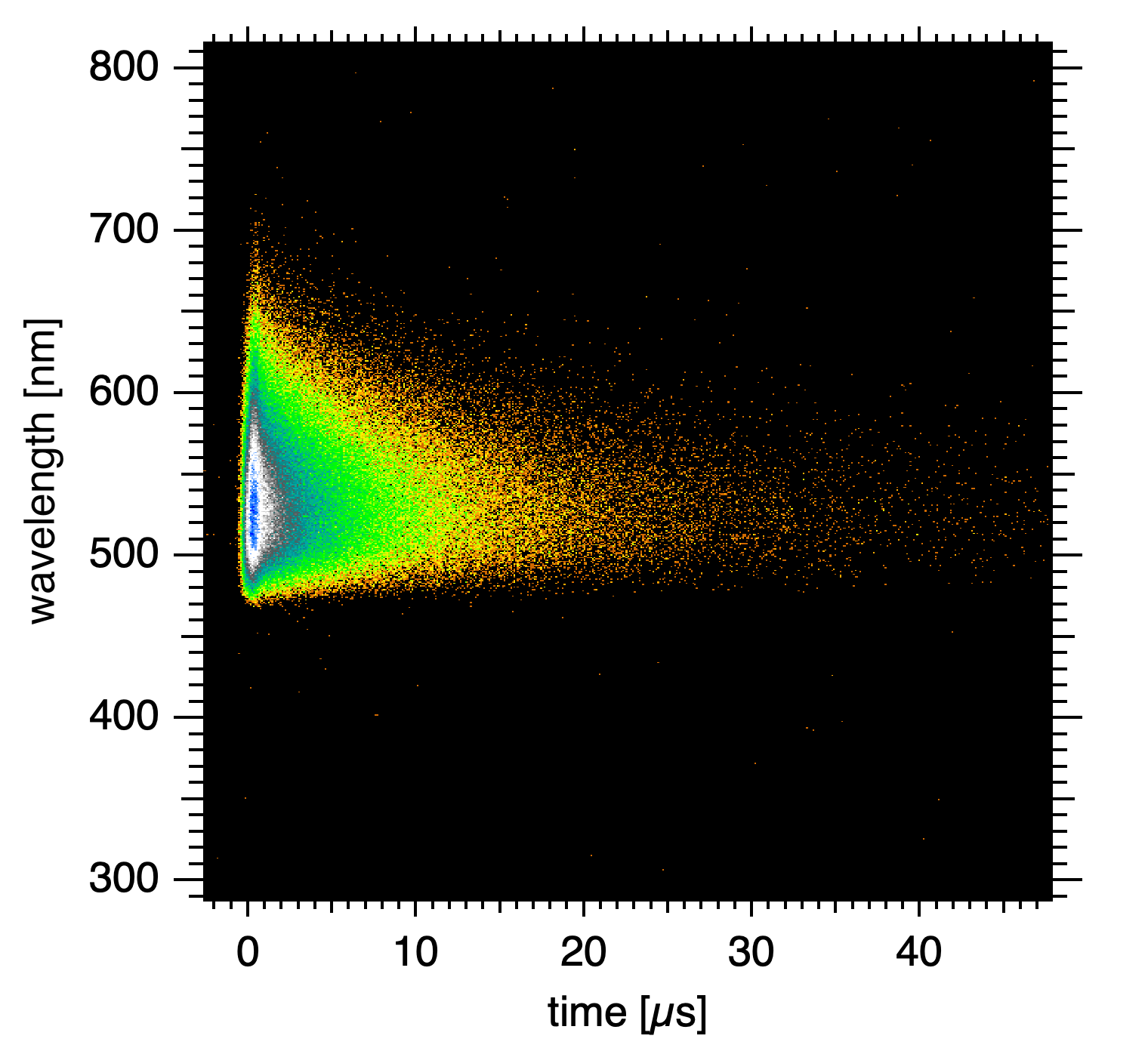}}

\subfigure[$\mathrm{4CzIPN}$ in PS]{\includegraphics*[scale=0.6]{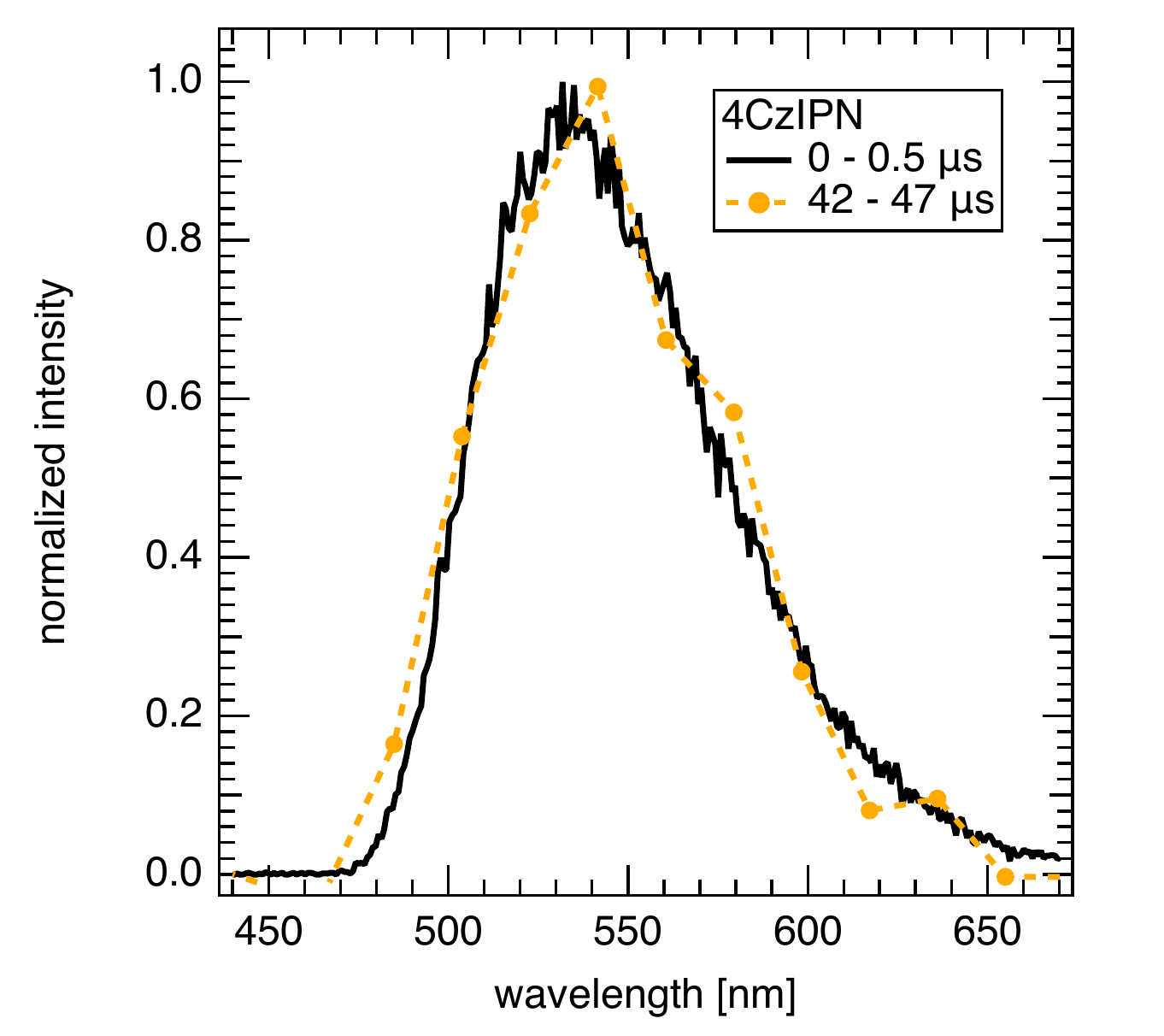}}
\qquad
\subfigure[$\mathrm{3CzClIPN}$ in PS]{\includegraphics*[scale=0.6]{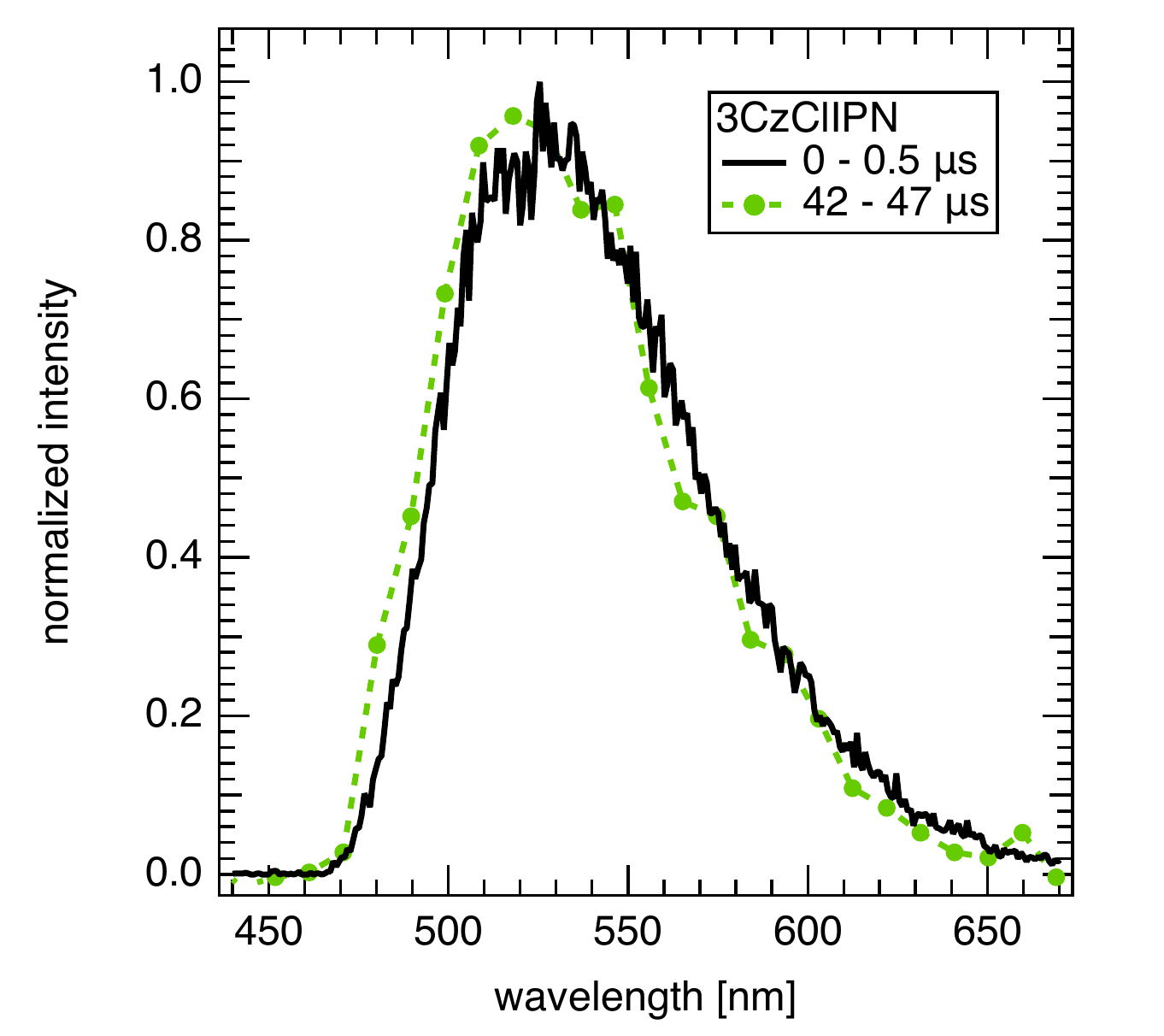}}
\caption{Time-resolved spectra}
\end{figure}

\noindent Streak measurements were performed on a Hamamatsu C10910-05 camera. Films were excited at 405\,nm with a pulse rate of 10\,kHz (Light Conversion, Pharos laser with Orpheus OPA) in an evacuated cryostate (JANIS, ST-500). Delayed spectra were binned to 12 data points. Note, that the PL part of the decay is distorted when measuring the full decay in a 50\,\textmu s window, as the IRF is longer than the nanosecond decay. To evaluate transients, the PL must be additionally measured in a smaller time window (50--200\,ns) at the same pulse rate. 

% \clearpage
% \section{OLED devices}
% \begin{figure}[h]
% \centering
% \subfigure[$J-V$ characteristics]{\includegraphics*[scale=0.6]{OLED_JV.pdf}}
% \qquad
% \subfigure[luminous flux]{\includegraphics*[scale=0.6]{OLED_flux.pdf}}
% \subfigure[power efficiency]{\includegraphics*[scale=0.6]{OLED_PEvsU.pdf}}
% \qquad
% \end{figure}
%The resulting devices were measured in an integrating sphere (UMBK-150, Gigahertz-Optik) with a photometric detector (VL-1101-1, Gigahertz-Optik) and a sourcemeter (Keithley, 2636B).

\clearpage

\section{Quantum chemical calculations}

\noindent Structural models of $\mathrm{4CzIPN}$ and 3CzIPN were generated with GaussView.\cite{GV} Their ground state geometries were optimized with density functional theory (DFT) calculations utilizing the B3LYP functional,\cite{Becke1988, Lee1988, Becke1993} which has been applied in previous studies of TADF emitter materials.\cite{Liang2016, DeSilva2019, Wang2019, Zhang2019, Montanaro2019, Weissenseel2019, Izumi2020} Furthermore, Grimme dispersion correction with Becke-Johnson damping (D3BJ)\cite{Grimme2010, Grimme2011} was employed to improve the description of the interaction between the carbazole units.\cite{Moral2015, Noguchi2017} During the optimizations, the double-\(\zeta\) Pople style basis set 6-31+G* was utilized.\cite{Hehre1977} Calculations of vibrational frequencies on the same level of theory were realized to check that the optimized structures correspond to minima on the potential energy surface. The subsequent time-dependent DFT (TD-DFT) calculations with and without Tamm-Dancoff approximation (TDA) for the first 20 excited states with singlet and triplet multiplicity were performed with the same basis set employing either the B3LYP functional as for the structure optimizations or the range-separated hybrid functional CAM-B3LYP.\cite{Yanai2004} The described quantum chemical calculations were realized with the Gaussian 16 software package.\cite{Gaussian} 

\noindent Furthermore, the Orca program\cite{ORCA, Neese2018} was used for the sTD-DFT and sTDA calculations\cite{Grimme2013, Bannwarth2014, Risthaus2014} with the same functionals and the somewhat larger basis set def2-SVPD.\cite{Rappoport2010} In these calculations, all excited states up to \(10 \, \mathrm{eV}\) were determined.\cite{Risthaus2014,Rappoport2010} In addition, Turbomole\cite{Furche2014, Turbomole} was employed for the excited state calculations with the approximate coupled cluster singles and doubles (CC2)\cite{Christiansen1996, Haettig2000, Haettig2002} as well as the algebraic diagrammatic construction to second order methods (ADC(2)).\cite{Trofimov1995, Haettig2005} Besides the unscaled methods, also the spin-component scaled (SCS) variants\cite{Grimme2003, Hellweg2008} were applied. These calculations utilized also the def2-SVPD basis set\cite{Rappoport2010} in combination with the corresponding auxiliary basis set for the resolution-of-identity (RI) approximation\cite{Hellweg2015} and an SCF convergence criterion of \(10^{-8} \, \mathrm{Hartree}\). Furthermore, core orbitals were frozen and only the first three excited states of singlet and triplet multiplicity were determined. 

\noindent Finally wavefunction analysis was performed with the TheoDORE software package.\cite{Plasser2012, TheoDORE} The obtained natural transition orbitals (NTOs) were visualized with Jmol.\cite{Jmol}

\section{Further results from excited state calculations}

\begin{table} [htbp]
\begin{tabular}{lcllcll} 

		&	$\mathrm{4CzIPN}$ &&&  	$\mathrm{3CzClIPN}$ & & \\

	&	$\Delta E_{\mathrm{ST}}$ & $S_1$ & $T_1$  & $\Delta E_{\mathrm{ST}}$ &$S_1$ & $T_1$ \\
	method &	$[\mathrm{meV}]$ & $[\mathrm{eV}]$ & $[\mathrm{eV}]$ &$[\mathrm{meV}]$ & $[\mathrm{eV}]$ & $[\mathrm{eV}]$ \\
\hline
\hline
\textbf{B3LYP} 	&&&&&&\\
TD-DFT 		&146 & 2.558 & 2.412        &103 & 2.460 & 2.356\\
TDA 			&154 & 2.578 & 2.423        &108 & 2.473 & 2.365\\
sTD-DFT 		&46   & 2.411 & 2.365        &41   & 2.341 & 2.300\\
sTDA 		&56   & 2.421 & 2.365        &47   & 2.347 & 2.300\\
\hline
\textbf{CAM-B3LYP} &&&&&&\\
TD-DFT 		&693 & 3.324 & 2.631         &589 & 3.259 & 2.670\\
TDA 			&359 & 3.359 & 3.000         &232 & 3.279 & 3.047\\	
sTD-DFT 		&77   & 3.056 & 2.979         &49   & 3.013 & 2.964\\
sTDA 		&87   & 3.067 & 2.980         &54   & 3.019 & 2.965\\
\hline
\textbf{Post-HF} &&&&&&\\
ADC(2)             &45 &2.765  &2.719         &20   & 2.770 & 2.751\\
SCS-ADC(2)     &42 &3.115   &3.072         &15   & 3.137 & 3.122\\
CC2                 &43 &2.842   &2.798         &22   & 2.847 & 2.824\\
SCS-CC2          &38 &3.179  &3.140         &16   & 3.204 & 3.188
\end{tabular}
\caption{Vertical singlet-triplet gaps and energies of the first excited singlet and triplet states relative to the optimized ground state.}
\label{tab:excitation}
\end{table}

\noindent To further assess the sTD-DFT and sTDA as well as the (SCS-)CC2 results, we performed additional calculations \textit{via} regular TD-DFT with and without TDA and \textit{via} ADC(2) with and without spin-component scaling. As summarized in table \ref{tab:excitation}, the computed values for the vertical energy gap between \(T_{1}\) and \(S_{1}\) ($\Delta E_{\mathrm{ST}}$) strongly depend on the choice of the functional in TD-DFT. The global hybrid functional B3LYP with 20\,\% exact exchange overestimates this splitting by a factor of approximately 2 relative to the experimental values, whereas the range-separated CAM-B3LYP functional yields significantly larger values. Our results for $\mathrm{4CzIPN}$ corroborate the findings that the choice of functional shows a pronounced effect on $\Delta E_{\mathrm{ST}}$ and that the results from the employed global hybrid functional are in better agreement with experiments than from the range-separated one, which have first been reported by Adachi \textit{et al.}\cite{uoyama2012highly}

\noindent Furthermore, the application of the TDA in the excited state calculations improves the results only in case of CAM-B3LYP, but the computed values for the vertical  $\Delta E_{\mathrm{ST}}$ are still much larger than with the B3LYP functional. For the latter, TDA does not lead to significant improvements. Relative to the results obtained with regular TD-DFT, the semiempirical sTD-DFT and sTDA approaches proposed by Grimme and coworkers\cite{Grimme2013, Bannwarth2014, Risthaus2014} yield  values of $\Delta E_{\mathrm{ST}}$, which are in better agreement with the Post-Hartree-Fock methods. Also these values do not exhibit such a pronounced dependence on the chosen functional as for the regular TD-DFT/TDA calculations, but the individual energies of the \(S_{1}\) and \(T_{1}\) states behave more similar.

\noindent The origin of the good performance of the semiempirical sTD-DFT and sTDA methods for $\Delta E_{\mathrm{ST}}$ might be traced back to the fact that they are based on regular DFT ground state calculations, but the parameters employed in the excited state calculations were fitted to reproduce excitation energies obtained from SCS-CC2 calculations.\cite{Risthaus2014} To assess the performance and reliability of this and related Post-Hartree-Fock methods, we performed corresponding calculations for $\mathrm{4CzIPN}$ and $\mathrm{3CzClIPN}$. For this purpose, we employed the two approximate second order methods CC2 and ADC(2) with and without SCS. All four methods result in a similar value for the difference of $\Delta E_{\mathrm{ST}}$ between $\mathrm{4CzIPN}$ and $\mathrm{3CzClIPN}$ ranging from \(21\) to \(27 \, \mathrm{meV}\). Furthermore, the individual values are below \(50 \, \mathrm{meV}\). In contrast to this, the vertical excitation energies of \(S_{1}\) and \(T_{1}\) exhibit a larger variance of around \(0.4 \, \mathrm{eV}\), which is reminiscent of the functional dependence in case of the DFT-based calculations. Overall, the employed second order methods result in a consistent picture for \(\Delta E_{\mathrm{ST}}\).

\section{Population analysis}

\noindent To better understand the increase in charge transfer character from $\mathrm{4CzIPN}$ to $\mathrm{3CzClIPN}$, the individual populations of the excited electron and hole for the three fragments can be analyzed for each state, see table \ref{tab:population}. The hole population on fragment 1 (\(h_{1}\)) and the electron population on fragment 2 (\(e_{2}\)) increase, whereas the hole population on fragment 2 (\(h_{2}\)) and electron population on fragment 1 (\(e_{1}\)) decrease. Therefore, the hole becomes more localized on fragment 1 and the electron on fragment 2. In contrast to this, the populations on fragment 3 remain nearly constant. Therefore, also this analysis shows that the localization of the excited electron and of the hole on different fragments is favored for $\mathrm{3CzClIPN}$ relative to $\mathrm{4CzIPN}$ resulting in a smaller overlap of the corresponding wavefunctions and a decrease of $\Delta E_{\mathrm{ST}}$.

\begin{table} [htbp]
\begin{tabular}{|l|l|l|l|l|} 
\hline
		&	$\mathrm{4CzIPN}$ &&  	$\mathrm{3CzClIPN}$ &  \\
\hline
 &	$T_{1}$ & $S_{1}$ & $T_{1} $ & $S_{1}$ \\
\hline
$h_{1}$ 		& 0.739 & 0.798 & 0.822 & 0.848 \\
$e_{1}$ 		& 0.164 & 0.165 & 0.113 & 0.113 \\
$h_{2}$ 		& 0.207 & 0.146 & 0.123 & 0.103 \\
$e_{2}$ 		& 0.620 & 0.618 & 0.664 & 0.665 \\
$h_{3}$ 		& 0.008 & 0.008 & 0.010 & 0.004 \\
$e_{3}$ 		& 0.170 & 0.169 & 0.177 & 0.178 \\
\hline
\end{tabular}
\caption{Electron ($e$) and hole ($h$) populations of the first excited singlet states ($S_{1}$) and the first triplet state ($T_{1}$) divided into three fragments for the two investigated molecules obtained from the SCS-CC2 calculations. The first fragment consists of the carbazole units, the second one of the benzene ring and in case of $\mathrm{3CzClIPN}$ also of the Cl atom, and the third one of the cyano groups.}
\label{tab:population}
\end{table}

%\begin{table} [htbp]
%\begin{tabular}{|l|l|l|l|l|l|l|} 
%\hline
%		&	$\mathrm{4CzIPN}$ &&&  	$\mathrm{3CzClIPN}$ & & \\
%\hline
%fragment &	$T_{1}$ & $S_{1}$ & $T_{2}$ &$T_{1}$ & $S_{1}$ & $T_{2}$ \\
%\hline
%$h_{1}$ 		& 0.739 & 0.798 & 0.560 & 0.822 & 0.848 & 0.613 \\
%$e_{1}$ 		& 0.164 & 0.165 & 0.229 & 0.113 & 0.113 & 0.117 \\
%$h_{2}$ 		& 0.207 & 0.146 & 0.324 & 0.123 & 0.103 & 0.276 \\
%$e_{2}$ 		& 0.620 & 0.618 & 0.583 & 0.664 & 0.665 & 0.675 \\
%$h_{3}$ 		& 0.008 & 0.008 & 0.060 & 0.010 & 0.004 & 0.061 \\
%$e_{3}$ 		& 0.170 & 0.169 & 0.133 & 0.177 & 0.178 & 0.158 \\
%\hline
%\end{tabular}
%\label{sim_ST}
%\caption{Electron ($e$) and hole ($h$) populations of the first excited singlet states ($S_{1}$) and of the first two triplet states ($T_{1}$ and $T_{2}$) divided into three fragments for the two investigated molecules. The first fragment consists of the carbazole units, the second one of the benzene ring and in case of $\mathrm{3CzClIPN}$ also of the Cl atom, and the third one of the cyano groups.}
%\end{table}

\clearpage

%\bibliography{A_bibo}
